\documentclass[twocolumn]{aastex631}
\usepackage{array}
\usepackage{threeparttable}
\usepackage{booktabs}
\usepackage{array}
\usepackage{multirow}
\usepackage[version=4]{mhchem}

\begin{document}


\author[0000-0002-3034-8505]{Anastasia Triantafillides}
\email{anastasia.triantafillides@gmail.com}
\affiliation{Department of Astronomy, University of Wisconsin–Madison, Madison, WI 53703, USA}
\affiliation{Wisconsin Center for Origins Research, University of Wisconsin–Madison, 475 N. Charter Street, Madison, WI 53706, USA}

\author[0009-0000-2075-1109]{Jason S. Pero}
\affiliation{Department of Physics and Astronomy, University of Kansas, Lawrence, KS, USA}

\author[0000-0001-7714-7551]{Angelique Kahle}
\affiliation{Max-Planck-Institut f\"ur Astronomie, K\"onigstuhl 17, D-69117 Heidelberg, Germany}

\author[0000-0002-9539-4203]{Thomas Beatty}
\affiliation{Department of Astronomy, University of Wisconsin–Madison, Madison, WI 53703, USA}
\affiliation{Wisconsin Center for Origins Research, University of Wisconsin–Madison, 475 N. Charter Street, Madison, WI 53706, USA}

\author[0000-0003-2215-8485]{Renyu Hu} 
\affiliation{Department of Physics and Astronomy, Rutgers, The State University of New Jersey, Piscataway, NJ 08854, USA}

\author[0000-0002-4404-0456]{Caroline V. Morley}
\affiliation{Department of Astronomy, University of Texas at Austin, 2515 Speedway, Stop C1400, Austin, TX 78751, USA}

\author[0000-0001-9667-9449]{David J. Wilson}
\affil{Laboratory for Atmospheric and Space Physics, University of Colorado, 600 UCB, Boulder, CO 80309}

\author[0000-0003-0514-1147]{Laura Kreidberg}
\affiliation{Max-Planck-Institut f\"ur Astronomie, K\"onigstuhl 17, D-69117 Heidelberg, Germany}

\author[0000-0001-5727-4094]{Drake Deming}
\affiliation{Department of Astronomy, University of Maryland, 4296 Stadium Drive, College Park, MD 20742, USA}

\author[0000-0002-1835-1891]{Ian J.M.\ Crossfield}
\affiliation{Department of Physics and Astronomy, University of Kansas, Lawrence, KS, USA}

\author[0009-0006-3846-4558]{Madison G. Scott}
\affiliation{School of Physics and Astronomy, University of Birmingham, Edgbaston, Birmingham B15 2TT, UK}

\title{The SPACE Program IV: Atmospheric Features on the Cold Transiting Sub-Neptune TOI-6478 b}

\begin{abstract}

Among the current population of exoplanets there are very few sub-Neptunes that have their atmospheric composition identified. This is primarily due to sub-Neptunes having high metallicity, clouds, and/or hazes obscuring absorption features. This often produces a spectrum that presents itself as a featureless spectrum especially in the near-IR. However, there are tentative trends that suggest sub-Neptunes with the lowest equilibrium temperatures ($\mathrm{T_{eq} < 400\:K}$) have detectable features, perhaps due to inefficient haze production or cloud sedimentation. TOI-6478 b is a low-density sub-Neptune exoplanet ($\rm <\mathrm{9.9\:M_{\oplus}}$ and $\rm \sim 4.5R_{\oplus}$) with an equilibrium temperature $\mathrm{T_{eq} = 220 \:K}$ assuming zero albedo. In this study, we present the highest S/N spectral feature ($\sim$6) of any sub-Neptune detected using HST/WFC3. We detect this spectral feature using a robust analysis that incorporates the Transit Light Source Effect (TLSE) to rule out false detections due to stellar activity. With the narrow bandpass of HST we were unable to distinguish whether the spectral feature at 1.4$\mu$m is from H$_2$O or CH$_4$ absorption. Meanwhile, self-consistent atmospheric models provide physical motivation for CH$_4$ to be the dominant absorber given that H$_2$O should condense or cold trap at higher pressures. Through our analysis we determine that HST/WFC3 alone cannot distinguish between major absorbers. However, a broader wavelength coverage by JWST will be able to measure the atmospheric composition in detail, due to the high S/N features for this target.

\end{abstract}

\section{Background}\label{sec:bkg}

Sub-Neptune exoplanets ($\lesssim4\:\mathrm{R_{\oplus}}$ and $\lesssim20\,\mathrm{M_{\oplus}}$) are the most common planets in the transit catalog to date, comprising $\sim40\%$ of the current population \citep{exoplanetarchive}. Because Neptune-like exoplanets are so common, they have become frequent targets for atmospheric characterization with JWST and HST \citep[e.g.,][]{benneke2019water, tsiaras2019water, madhusudhan2023carbon, wogan2024jwst, ehrenreich2014near, benneke2019sub, berta2012flat, kreidberg2014clouds, kempton2023reflective, schlawin2024possible, ohno2025possible, mikal2023hubble, benneke2024jwst, holmberg2024possible,
hu2025water,knutson2014featureless, knutson2014hubble, guo2020updated, mikal2021transmission, kreidberg2022tentative,roy2023water, crossfield2020phase, dragomir2020spitzer, edwards2023characterizing, radica2024muted,fraine2014water, chachan2019hubble, murphy2025hst, wallack2024jwst, gordon2026jwst, wallack2026jwst, meech2026jwst}. However, despite how prolific observations of Neptune-like atmospheres has been, characterizing their atmospheres has proven difficult because of a lack of spectral clarity due to a combination of high metallicity, clouds, and/or hazes \citep[e.g.,][]{kreidberg2014clouds, kempton2023reflective, knutson2014featureless, knutson2014hubble, libby2020featureless, wallack2024jwst, gordon2026jwst, wallack2026jwst, meech2026jwst, libby2026james}. Prior to launch, one of JWST's primary science goals was to deliver on measured abundances of major carbon- and oxygen-bearing species for Neptune-like atmospheres, both to better understand their formation mechanisms and to contextualize the formation of gas giants in the Solar System. However, because many atmospheric observations of Neptune-like exoplanets have yielded muted transmission spectra, the formation pathways of Neptune-like planets close to their host stars has remained elusive \citep{bean2021nature}. 

Currently, there are still few measurements of carbon- and oxygen-bearing species needed to derive the C/H and O/H ratios used to infer the formation histories of Neptune-like exoplanets \citep{oberg2011effects}. To date, abundances of H$_2$O, CO$_2$, and CH$_4$ (the major carbon- and oxygen-bearing species at cool temperatures) have been measured for only a handful of sub-Neptunes. Both TOI-270 d \citep{benneke2024jwst, holmberg2024possible} and GJ 3470 b \citep{beatty2024sulfur} have detections of all three molecules, whereas K2-18 b \citep{hu2025water, madhusudhan2023carbon} and TOI-1231 b \citep{sarkar2026transmission} have detections of CO$_2$ and CH$_4$ with implications of H$_2$O at higher pressures. TOI-732 c has only a CH$_4$ detection \citep{rigby2025jwst}, and the transmission spectrum of TOI-260 b provides an upper limit on CH$_4$ \citep{meech2026jwst}. In the Solar System's giant planets the dominant carbon-bearing molecule is CH$_4$ \citep{irwin09}, and the overall carbon abundance increases with decreasing planet mass \citep{madhusudhan2025exploring}. The individual molecular abundances for of TOI-270 d fall below this solar system trend, while K2-18 b, GJ 3470 b, TOI-732 c, and TOI-1231 b remain consistent with it \citep{madhusudhan2025exploring, sarkar2026transmission}. With only four planets that have definitive detections, it is not yet clear whether Neptune-like exoplanets overall follow this Solar System trend. Testing it, and using molecular abundances to identify how Neptune-like planets form, requires planets whose atmospheres are clear enough to show spectral features.

Which sub-Neptunes have clear atmospheres appears to depend on temperature \citep{yu2021haze, brande2024clouds}. \cite{brande2024clouds} identified a trend in sub-Neptune cloudiness with equilibrium temperature: atmospheric clarity increases significantly for the coldest ($<500\mathrm{\:K}$) and hottest ($>700\mathrm{\:K}$) planets, with strong aerosol attenuation at intermediate temperatures. Their models indicate that the observed population is best described by a low cloud sedimentation efficiency ($f_\mathrm{sed}\sim 0.1$) and a high metallicity ($100\times$ solar), and can be used to predict which planets will have clouds that obscure gas absorption in space-based observations. A low sedimentation efficiency typically results in smaller particles in a more lofted cloud layer, whereas a higher sedimentation efficiency results in larger cloud particles deeper in the atmosphere \citep{morley2015thermal}.

There are a few possible mechanisms for the clearer atmospheres at the colder end of this trend. One being the growth of large cloud particles, which then sink below the photosphere. A similar lowering of the cloud deck has been proposed for cooler brown dwarfs, in which increasing cloud particle sizes lead to rapid rainout that thins the clouds \citep{tsuji2003transition, knapp2004near}. The same mechanism may operate in temperate sub-Neptunes with $\mathrm{T_{eq} < 300\:K}$ if their sedimentation efficiency is high. At these low equilibrium temperatures, large cloud particles would form a vertically compact layer and sink below the photosphere, allowing the gas species in the upper atmosphere to be probed directly with transit spectroscopy \citep{morley2015thermal}. Alternatively, for the case of a hazy atmosphere, it has been hypothesized that cooler habitable zone exoplanets ($<400\mathrm{\:K}$) may be less hazy as a result of possessing high-surface-energy hazes that can be removed efficiently \citep{yu2021haze}.

However, \cite{brande2024clouds} noted that their observational sample lacked a significant number of sub-Neptunes with $\mathrm{T_{eq}<500\:K}$, despite the expanded sample size compared to previous works \citep[e.g.,][]{crossfield:2017b}. Obtaining more atmospheric observations of temperate planets is therefore a priority, both to anchor the lower-temperature end of the observed trend and to confirm the cloud physics predicted by atmospheric models. This need is especially urgent for JWST observations. Because they are predicted to have clear atmospheres, temperate sub-Neptunes are also promising targets for measuring carbon- and oxygen-bearing species. At these low temperatures, H$_2$O and CH$_4$ can produce overlapping absorption bands in the HST/WFC3 bandpass, depending on their relative abundances, which makes them difficult to disentangle with WFC3 alone. JWST can break this degeneracy, as has been shown for other sub-Neptunes \citep{benneke2019sub, madhusudhan2023carbon, murphy2025hst}, and an HST/WFC3 spectrum can still show whether an atmosphere is clear enough for such follow-up.

TOI-6478\,b is a recently discovered Neptune-like exoplanet that orbits an M5-type star with a period of $\sim34$ days, at a scaled semi-major axis of $a/R_* \sim 104$, and has a 3$\sigma$ upper mass limit of $\mathrm{9.9\:M_{\oplus}}$ \citep{scott2025toi}. These orbital and stellar parameters make TOI-6478 b the coldest transiting sub-Neptune detected to date ($\mathrm{T_{eq} = 220 \:K}$ assuming zero albedo), and its mass limit implies a surprisingly low density for a temperate sub-Neptune. Here, we present HST/WFC3 transmission spectrum observations of TOI-6478 b.

\section{Observation, Data Reduction, and Lightcurve Fitting}\label{sec:obs_methods}
We observed two transits of TOI-6478\,b using Wide Field Camera 3 (WFC3) on the Hubble Space Telescope (HST) as a part of the SPACE\footnote{Sub-Neptune Planetary Atmosphere Characterization Experiment} Program (PIDs 17192 \& 17414, co-PIs: Laura Kreidberg and Drake Deming), a multi-cycle HST program designed to survey the atmospheres of sub-Neptunes. All the HST data used in this paper can be found in MAST: \dataset[10.17909/y21f-6a73]{https://doi.org/10.17909/y21f-6a73}. The first transit we observed occurred on UT 2025 Nov. 5, and the second occurred on UT 2025 Dec. 9. We observed both transits using the G141 grism in spatial scanning mode and used five HST orbits of observing time for each visit. At the start of each orbit we took a single direct image using the F139M filter to exactly determine the pixel position of the star on the detector. The G141 exposures used the RAPID sampling mode with 7 samples up the ramp, giving an individual exposure time of 112 seconds. We scanned the spectra on the detector using round-trip scanning at a rate of 0''.031 per second, giving a total scan height of approximately 26 pixels. 



\subsubsection{STIS Stellar UV Spectrum}
\begin{figure}
    \centering
    \includegraphics[width=\linewidth]{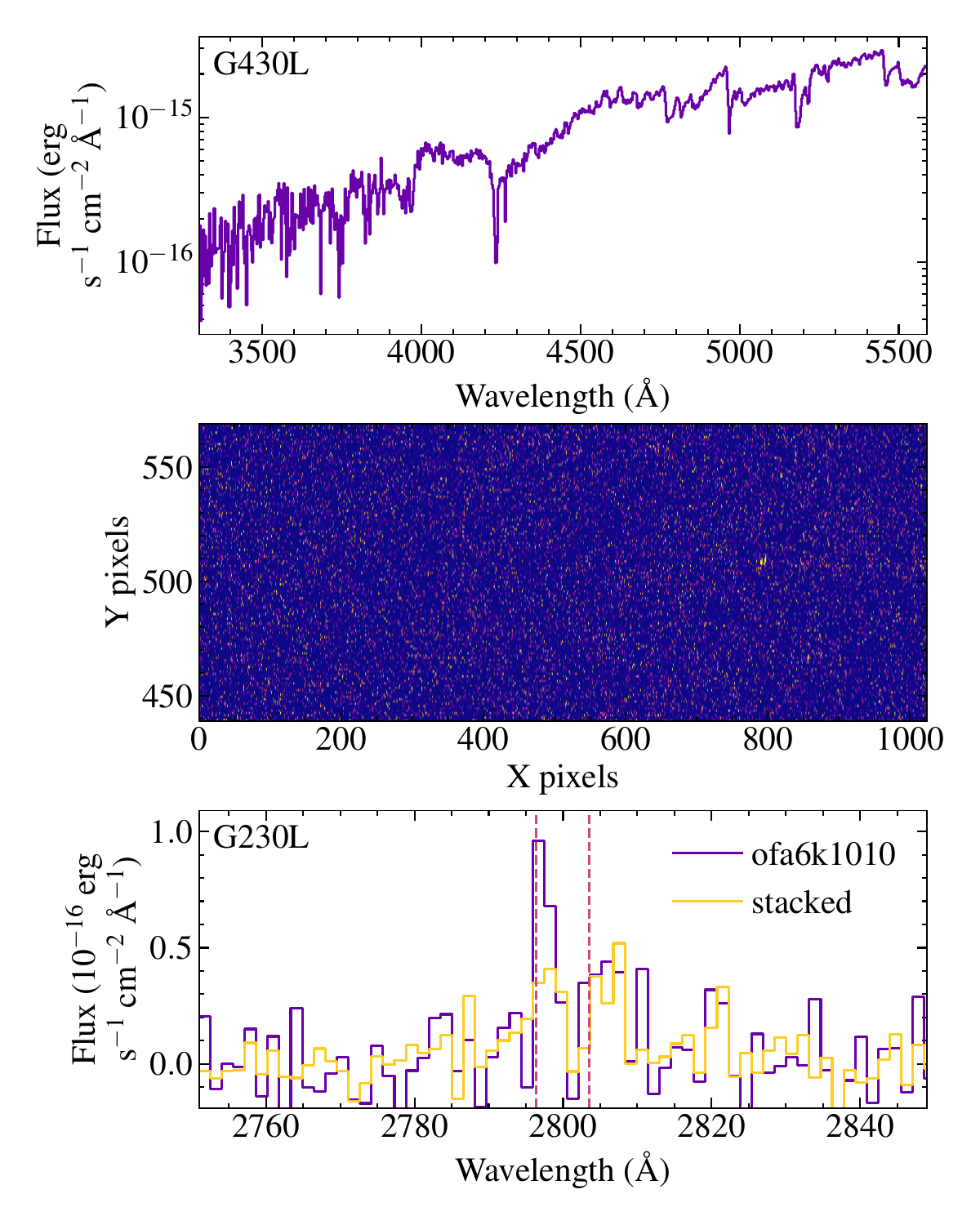}
    \caption{STIS spectroscopy of TOI-6478. Top: The G430L spectrum. Middle: Section of the stacked G230L flt images, showing the \ion{Mg}{2}\,2800\AA\ lines but no other features. Bottom: Comparison of G230L spectra around \ion{Mg}{2} extracted from just the first observation and from the stacked flt image. }
    \label{fig:stis}
\end{figure}

In addition to the WFC3 transits, we also obtained ultraviolet spectroscopy of the TOI-6478 host star using the Space Telescope Imaging Spectrograph \citep[STIS][]{woodgateetal98-1}also on HST. STIS observations were obtained on 2025~December~24 and 2025~December~27. Two exposures used the G430L grating covering $\approx2900-5700$\,\AA\ with a total exposure time of 1880\,s, and a further five exposures used the G230L grating covering $\approx1570–3180$\,\AA\ and a total exposure time of 8710\,s.

For the G430L exposure we retrieved and inspected the automated Hubble Advanced Spectral Project \citep[HASP][]{debesetal24-1} coadd of the two exposures. The target is detected with a signal to noise ratio  of 10--100 across essentially the entire bandpass (Figure \ref{fig:stis}, top). 

Conversely, the star is barely detected in the G230L data, with the only detected feature being the \ion{Mg}{2}\,2800\,\AA\ doublet which is visible by eye in the flt images. We stacked the five G230L flt images into one (Figure \ref{fig:stis}, middle) and extracted a spectrum with {\sc stistools}\footnote{\url{https://stistools.readthedocs.io/en/latest/index.html}}, fixing the trace location to the observed \ion{Mg}{2} lines. Even in the stacked image no features outside of \ion{Mg}{2} are detected. We found a integrated, background subtracted flux for the \ion{Mg}{2} lines of $(4.0\pm 0.5)\times10^{-16}$\,erg\,s$^{-1}$\,cm$^{-2}$. Intriguingly, the first exposure (ofa6k1010) appears brighter in both the flt image and in the extracted spectrum (Figure \ref{fig:stis}, bottom). This could be due to a small flare, but the signal is too small for reliable further analysis. 

\subsection{Data Reduction}

To ensure our results are robust, we performed two separate reductions of the HST/WFC3 spectra using the \texttt{PACMAN} pipeline \citep{pacman}. Here we briefly describe the data reduction process; \cite{pacman} provides a more detailed description of all of the \texttt{PACMAN} data reduction and fitting stages. Both reductions share the same initial Stage 00-03 processing and final wavelength binning and differ only in the spectral-trace extraction and background subtraction used for the secondary analysis. The primary analysis used \texttt{Pegasus}\footnote{https://github.com/TGBeatty/PegasusProject} to fit the reduced broadband and spectroscopic light curves, while our secondary analysis used the lightcurve fitting routines built into \texttt{PACMAN}. Figure \ref{fig:reduction comparison} shows the results of both reductions for comparison.

For the primary analysis, we began with the \texttt{\_ima} files and background-corrected each difference image by subtracting the median value of all pixels with counts below 800. Additionally, both visits include stray light from the companion star LP 789-76 on the lower edge of the detector. For the secondary analysis, we avoid a bias to our background flux by estimating it in a box between rows 201-261 and columns 2-261 on the upper edge of the detector. For background subtraction of the secondary analysis, we use the median flux of all pixels in this box. 

We then create a reference spectrum based on the throughput of G141 and the \texttt{PHOENIX} stellar model grids \citep{allard2011stellar} with the stellar parameters $T_{\mathrm{eff}} = 3230\:\mathrm{K}$, $\log(g) = 5.016$, and metallicity = $-0.53$ from \cite{scott2025toi}. We determine the position of the star in the direct images to create an initial guess for the wavelength solution using the known dispersion of the grism. Using the reference spectrum as a template, we determine a shift and scaling in wavelength-space that minimizes the difference between the template and the first spectrum in the visit. This first exposure in the visit is then used as the template for the following exposures in the visit. The secondary analysis used a separate spectral-trace extraction in \texttt{PACMAN}'s stage 10. The observed field contains many faint secondary sources, which appear as faint traces on the detector. To avoid an imprint of these traces on the planetary spectrum, we set the extraction region such that it excludes them in secondary analysis. 

We then use the optimal spectral extraction outlined in \cite{horne1986optimal} with a 9 pixel smoothing and a 15 sigma cut to mask cosmic rays. Finally, we created a broadband lightcurve by summing between the full wavelength range 1.160 - 1.617 $\mu$m and spectroscopic lightcurves by summing in evenly spaced bins of $\Delta\lambda= 0.024\mu\mathrm{m}$ resulting in 20 bins total.

\subsection{Lightcurve fitting}\label{sec:lightcurve_fitting}

For both analyses, we fit the broadband and spectroscopic lightcurves using the \texttt{BATMAN} transit model \citep{kreidberg2015batman}. To start, we simultaneously fit the broadband lightcurves of both visits to better constrain the orbital and system parameters for the spectroscopic fits. We did so assuming a circular orbit, and fit both visits using a single set of transit parameters and a single radius ratio ($R_p/R_*$). We imposed Gaussian priors on the period (P), time of conjunction ($\mathrm{T_c}$), cosine of the inclination ($\cos i$), and the log of the scaled semi-major axis ($\log(a/R_*)$) based on the measurements made by \cite{scott2025toi}. We did not impose a prior on $R_p/R_*$. Additionally, we used a quadratic limb-darkening law, and fixed the limb-darkening coefficients in this broadband fit to the values generated for the star TOI-6478 using \texttt{ExoTic-LD} \citep{Grant2024} and the MPS1 stellar models \citep{kostogryz2023mps}. In addition to the astrophysical parameters, for each visit we also fit for a background, visit-long, linear trend in time -- and orbit-long exponential hooks. For the primary analysis, we discarded the observations from the first orbit in each visit, and used a separate hook model for the second orbit of each set of observations (the ``first'' orbit of the used data). Prior to fitting the spectroscopic data in the primary analysis, we divided the exponential ramps of the broadband lightcurves from the each spectroscopic channel.


Next, for both analyses, we performed the spectroscopic fits and sampled the posteriors using the \texttt{emcee} \citep{foreman2013emcee} Python package. For the primary reduction, we sampled the parameter space with 40 walkers, 4 chains, and a 20,000-step run with 2000-step burn-in and for the secondary reduction, we sampled with 100 walkers and a 30000-step chain, removing the first 3000 steps as burn-in. For this fit, we fixed the orbital parameters to the results from the joint broadband fit. We then only fit a scaled planetary radius for both visits simultaneously with separate parameters for a linear trend in time for each spectroscopic channel. Similar to the broadband fit, we kept the limb darkening parameters fixed using \texttt{ExoTic-LD} and the MPS1 stellar model \citep{Grant2024,kostogryz2023mps}. Since M-dwarfs have water in their photosphere, this can shows up in the limb darkening for the spectroscopic fits. To address this, we did an additional fit where we added the limb darkening parameters as a free variable. This resulting spectrum where we kept limb darkening as a free parameter is consistent with the nominal spectrum where we kept the limb darkening fixed. For the primary analysis, we inflated the uncertainties by applying a correction factor post-fit to avoid an underestimation in the transit depth uncertainties. We calculated this correction factor using the standard deviation of the residuals divided by the mean error. We then multiplied this correction factor to the uncertainties of the scaled radii values. For the secondary analysis, we fit for the parameter that inflates the uncertainties directly. Finally, for the primary analysis we confirmed that the fit had converged by ensuring the Gelman-Rubin statistic was below 1.1 for all parameters and for the secondary analysis, the chains of all parameters exceed 50 times the autocorrelation time, verifying the convergence of the sampling.

\subsection{Comparison of the Two Data Analyses}

\begin{figure}[h!]
    \centering
    \includegraphics[scale=0.28]{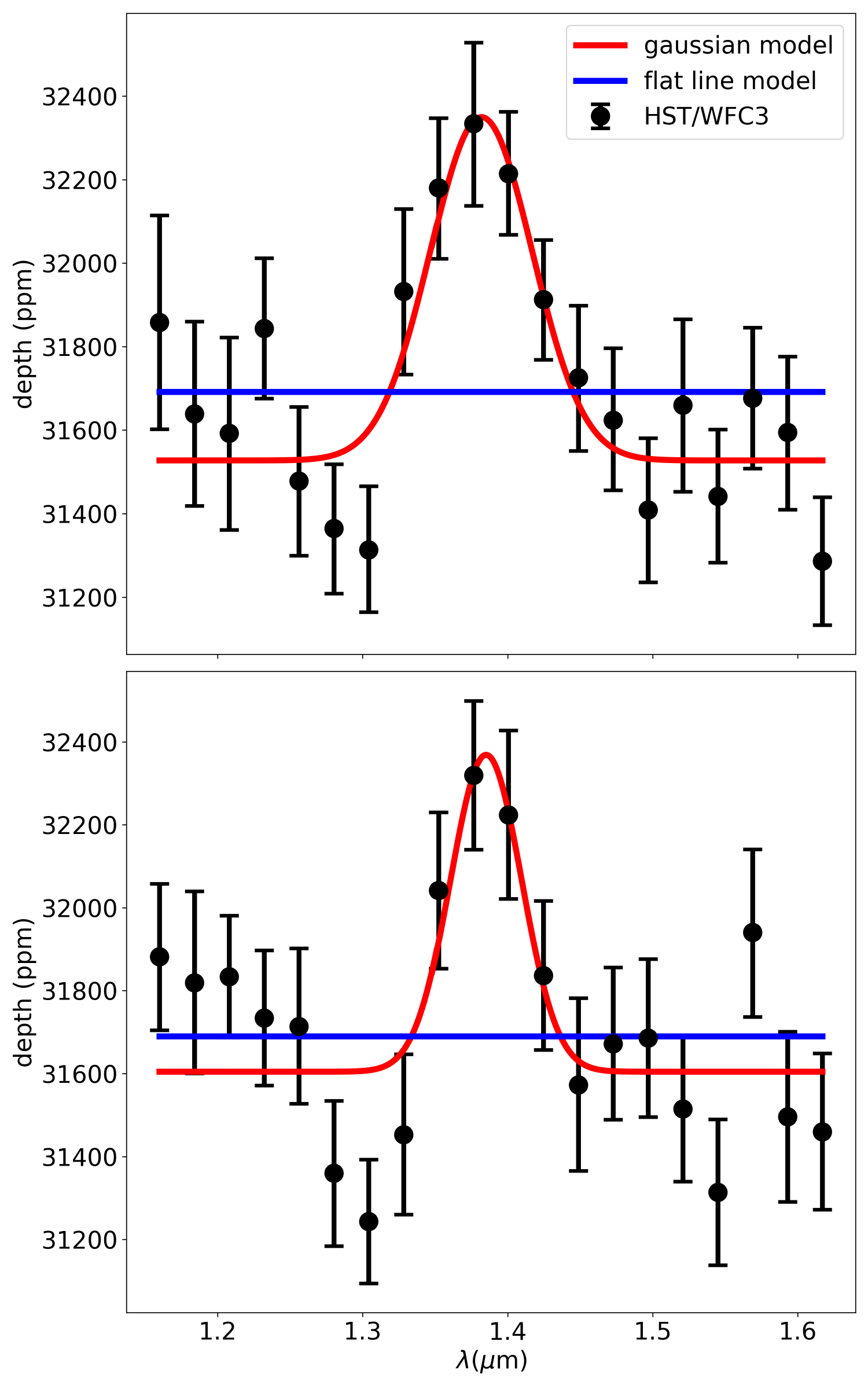}
    \caption{Top panel: Primary analysis in black with one sigma error bars and it's corresponding Gaussian model in red and flat line in blue. Bottom panel: same as top panel but for the secondary analysis. We find that the Gaussian model was preferred for both analyses with a $\Delta\ln{\mathrm{B}} = 22.8$ and $\Delta\ln{\mathrm{B}} = 17.28$ for the primary and secondary analyses respectively. We also find that they are consistent with one another by calculating a Bayes factor between the primary and secondary analyses of $\Delta\ln{\mathrm{B}} = 3.51$.}
    \label{fig:reduction comparison}
\end{figure}

To show that there is a significant feature in the HST/WFC3 spectrum, we fit a four-parameter Gaussian feature (constant offset, center time, width and amplitude) and a single-parameter flat line to both reductions, and compared the Bayesian evidence to estimate which model was preferred by the data. We find that the Gaussian model was preferred for the primary analysis at $\Delta\ln{\mathrm{B}} = 22.8$ and for the secondary analysis at $\Delta\ln{\mathrm{B}} = 17.28$, indicating that the 1.4$\mu$m feature is significantly present in the data itself. We also find that the two analyses are consistent with each other by calculating the Bayes evidence of their separate Gaussian fits. The difference in the resulting Bayes factor between the primary and secondary analyses is $\Delta\ln{\mathrm{B}} = 3.51$ demonstrating that the reductions are mildly, but not significantly different. In Figure \ref{fig:reduction comparison}, we show the two separate analyses and their corresponding Gaussian and flat line models. This model comparison demonstrates that when we use the the same system parameters, limb-darkening coefficients, and freely-fit the background systematics model, the two resulting transmission spectra for TOI-6478 b agree well as most wavelength bins agree within 1 sigma. 

\subsubsection{Using TESS data to check for stellar variability}

As a check, we compared the scatter in the TESS sector 35 and 62 lightcurves from \cite{scott2025toi} to assess whether the HST/WFC3 spectrum could be significantly contaminated by stellar variability. We did not use sectors 99 and 100 as the primary (LP789-76) and secondary (TOI-6478) stars were too close to disentangle. In \cite{scott2025toi} they isolated the primary (LP789-76) and secondary (TOI-6478) stars in the TESS full-frame images (FFIs) by taking a small pixel aperture around each stars location. The two isolated TESS light curves show no significant variability, so we conclude that the primary star does not significantly affect the HST/WFC3 spectrum.

Next, we used the formula for calculating observed Transit Light Source Effect (TLSE) from \cite{rackham2018transit} (Equation 1) to determine what spot-covering fraction would be required to match the observed spectral amplitude in the HST/WFC3 spectrum. For this calculation, it is typical to use a $\Delta \rm T_{eff} = 500 K$ for the different between the effective temperature of the stars photosphere and stellar heterogeneities \cite{rackham2018transit}. We used a PHOENIX stellar model with an effective temperature ($\mathrm{T_{eff}}$) of 3200 K as a proxy for the flux from TOI-6478's photosphere and a PHOENIX stellar model with an effective temperature ($\mathrm{T_{eff}}$) of 2700 K as a proxy for the flux from stellar heterogeneities. We find that the required spot covering fraction to match the spectral amplitude is $f_{\mathrm{het}}\sim 0.2$. Using the estimate for M5-type stars in Figure 4 from \cite{rackham2018transit} we estimated an expected peak-to-trough I-band variability amplitude of $\sim 0.013$ for solar-like spots and $\sim 0.04$ for giant spots. When we compare this to the scatter from TESS sectors 35 and 62 from \cite{scott2025toi} which is $\sim0.05$ we are able to rule out giant spots but are unable to rule out star spots from solar like spots. Further, in section \ref{sec:ExoTR}, we fit for the TLSE parameters and find that including the TLSE parameters does not significantly affect the results of our atmospheric retrievals. In order to definitively rule out stellar activity in this system we would need higher resolution photometric modeling. 

\section{Atmospheric modeling}\label{sec:results}

We used three complementary modeling frameworks to investigate the origin of the absorption features in TOI-6478b's WFC3 transmission spectrum. We first performed a set of free-chemistry retrievals with \texttt{petitRADTRANS} \citep{Molliere:2019prt, Blain:2024prt} and the Exoplanetary Transmission Retrieval framework \texttt{ExoTR} \citep{damiano2024lhs}, using different treatments of clouds and stellar heterogeneity, to assess which molecular absorption features were detected in the data. We then used self-consistent \texttt{EGP} forward models \citep{marley1999reflected, fortney2008synthetic, morley2015thermal} to evaluate these possible detections against the planet's expected atmospheric temperature structure and chemical composition.

Both retrieval analyses identified a significant molecular absorption feature near 1.4\,$\mu$m, but neither could uniquely identify if this feature was caused by H$_2$O or CH$_4$. The retrievals mildly favored H$_2$O over CH$_4$, but could not significantly distinguish between the two absorbers. This preference for water was initially surprising given the planets low equilibrium temperature. At such low temperatures, one would expect water to have condensed out of the atmosphere and for there to be little to no water vapor, a physical argument for methane as the source of the absorption.

The self-consistent forward models, however, indicate that water condensation does not necessarily eliminate all the observable water absorption. The models predict water-ice formation and a spectra dominated by CH$_4$ in the HST bandpass, however, it also contains a small contributions from water vapor at high altitudes. Thus, while condensation is expected to suppress water gas absorption at these low temperatures, it does not remove it entirely.

\subsection{Free-chemistry retrievals}\label{sec:retrievals}
\subsubsection{petitRADTRANS free retrieval}

We performed the free-chemistry atmospheric retrievals with the open source tool \texttt{petitRADTRANS} (pRT) version 3.2.0 (\citealt{Molliere:2019prt, Blain:2024prt}), using its retrieval framework as described in \cite{Nasedkin:2024prt}. We used \texttt{PyMultinest} \citep{Buchner:2014_PyMultinest}, a Python implementation of the nested sampling algorithm \texttt{MultiNest} \citep{Feroz:2009multinest}, to sample the parameter space. 

We specify each forward model by the pressure–temperature profile, planetary and stellar parameters, and the atmospheric opacity structure (i.e., composition). We first tested an initial Guillot \citep{Guillot:2010} pressure--temperature parameterization constrained the temperature over only a narrow pressure range; therefore, we adopted an isothermal profile with a uniform temperature prior of 100 to 300 K. We modeled the atmosphere in a pressure range between log$_{10}(P[\mathrm{bar}])=2)$ and log$_{10}(P[\mathrm{bar}])=-8)$.

\begin{deluxetable}{ll@{\hspace{0.20\columnwidth}}l}
\tablecaption{Retrieval Priors}
\label{tab:prt_retrieval_priors}
\setlength{\tabcolsep}{0pt}
\tablehead{
\multicolumn{1}{l}{Parameter} & & \colhead{Prior}
}
\startdata
\addlinespace[3pt]
$R\mathrm{_p}$ (R$_\oplus$) & & $\mathcal{N}(4.54, 0.27)$ \\
log$_{10}(g)$ & log$_{10}$(cm s$^{-2}$) & $\mathcal{U}(2.0, 3.1)$ \\
log$_{10}(P_\mathrm{ref}$) & log$_{10}$(bar) & $\mathcal{U}(-8.0, 2.0)$ \\ 
log$_{10}(P_\mathrm{cloud}$)\,\,\,\,\, & log$_{10}$(bar) & $\mathcal{U}(-6.0, 2.0)$ \\ 
T (K) & & $\mathcal{U}(100, 300)$ \\
H$_2$O & log$_{10}$(mass frac) & $\mathcal{U}(-12.0, 0.0)$ \\
CH$_4$ & log$_{10}$(mass frac) & $\mathcal{U}(-12.0, 0.0)$ \\
CO & log$_{10}$(mass frac) & $\mathcal{U}(-12.0, 0.0)$ \\
CO$_2$ & log$_{10}$(mass frac) & $\mathcal{U}(-12.0, 0.0)$ \\
NH$_3$ & log$_{10}$(mass frac) & $\mathcal{U}(-12.0, 0.0)$ \\
HCN & log$_{10}$(mass frac) & $\mathcal{U}(-12.0, 0.0)$ \\
$b$ & & $\mathcal{U}(-9.67, -5.18)$ \\
\addlinespace[3pt]
\enddata
\end{deluxetable}

We scaled the data uncertainties following the approach of \cite{Molliere:2025} (see also \citealt{Line:2015}). In this framework, we introduce an additional variance term by retrieving the uncertainty scaling parameter $b$, such that the scaled uncertainties are given by
\begin{equation}
\sigma_{\mathrm{scaled}}(\lambda) = \left[\sigma^2(\lambda) + 10^{b}\right]^{1/2},
\end{equation}
where $\sigma_\lambda$ are the observational uncertainties. We place a uniform prior on $b$, with lower bound $\log_{10}(0.01\min\left[\sigma(\lambda)\right]^2)$ and upper bound $\log_{10}(100\max\left[\sigma(\lambda)\right]^2)$.

We adopt a free chemistry framework in our retrieval analysis, including the following line opacities: \ce{CH4} \citep{Yurchenko:2024}, CO \citep{Rothman:2010}, \ce{CO2} \citep{Yurchenko:2020}, \ce{HCN} \citep{Barber:2014}, \ce{H2O} \citep{Polyansky:2018}, \ce{NH3} \citep{Coles:2019}. We selected these molecular species because their rovibrational transitions can produce observable spectral features within the observed wavelength range, making them suitable for atmospheric retrieval analysis. The selected wavelength range of 1.160-1.617 $\mu m$ contains strong \ce{H2O} and \ce{CH4} absorption features, which typically dominate the near-infrared opacity. The remaining molecules exhibit overlapping, yet weaker, features within this region.  We account for the presence or absence of clouds by parameterizing an opaque gray cloud deck with a uniform prior on the cloud-top pressure, \(\log_{10}(P_{\mathrm{cloud}}) \sim \mathcal{U}(-6.0, 2.0)\). 

We computed all forward models using correlated-$k$ opacities re-binned from $R=1000$ to $R=200$ with \texttt{Exo\_k} \citep{Leconte:2021}. We then ran Multinest to convergence with 1000 live points and a sampling efficiency of 80\%. 

\begin{deluxetable}{l@{\hspace{0.07\columnwidth}}cc}
\tablecaption{Bayesian Evidence for \texttt{petitRADTRANS} Retrieval Models}
\label{tab:prt_bayesian_evidence}
\setlength{\tabcolsep}{0pt}
\tablehead{
\multicolumn{1}{l}{Retrieval Model} & \colhead{ln(Z)} & \colhead{\hspace{15pt}$\Delta$ln(Z)}
}
\startdata
\addlinespace[3pt]
\ce{H2O}, \ce{CH4}, \ce{NH3}, \ce{CO}, \ce{CO2}, \ce{HCN}, clear & $-18.9$ & $0.0$ \\
\addlinespace[3pt]
\tableline
\addlinespace[3pt]
No \ce{H2O} & $-21.1$ & $2.2$ \\
No \ce{CO} & $-19.3$ & $0.4$ \\
No \ce{CH4} & $-18.5$ & $-0.4$ \\
No \ce{HCN} & $-18.5$ & $-0.4$ \\
No \ce{CO2} & $-17.5$ & $-1.4$ \\
No \ce{NH3} & $-17.3$ & $-1.6$ \\
\addlinespace[3pt]
\tableline
\addlinespace[3pt]
\ce{H2O}-only, clear & $-17.6$ & N/A \\
\addlinespace[3pt]
\ce{H2O}-only, cloud & $-18.2$ & N/A \\
\addlinespace[3pt]
\ce{CH4}-only, clear & $-20.3$ & N/A \\
\addlinespace[3pt]
\ce{CH4}-only, cloud & $-20.4$ & N/A \\
\addlinespace[3pt]
\enddata
\end{deluxetable}

We compare models using the difference in Bayesian evidence, \(\Delta \ln Z\), between the cloudy and clear-atmosphere cases. For both the full chemical model and the \ce{H2O}-only model, the clear-atmosphere models yield slightly higher evidence (i.e., where we defined \(\Delta \ln Z < 0\) as cloudy minus clear). However, the magnitude of \(\Delta \ln Z\) is small, indicating that the preference for a clear atmosphere is not statistically significant. 

\begin{figure}
\centering
\includegraphics[scale=0.5]{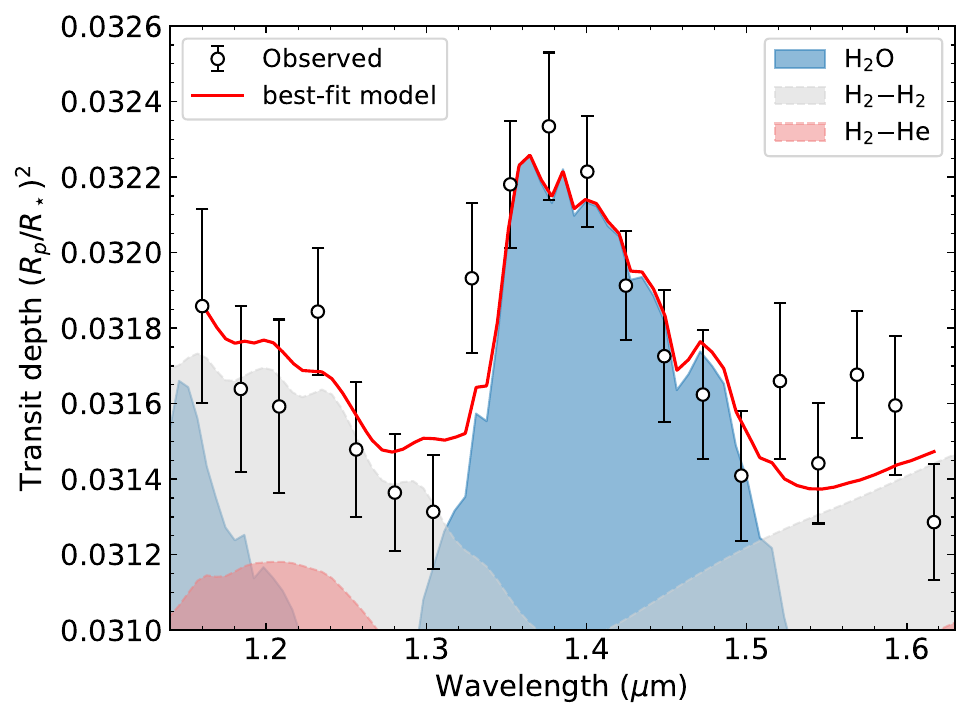}
\caption{Best-fit \ce{H2O}-only retrieval model of the HST transmission spectrum of TOI-6478 b. The wavelength-dependent opacity contributions from H$_2$O and the gas continuum are shown. \ce{H2}--\ce{H2} and \ce{H2}--He continuum opacity are indicated by the gray and red shaded regions, respectively.}
\label{fig:prt_retrieval_h2o_only_bestfit_spectrum}
\end{figure}

\begin{figure}
\centering
\includegraphics[scale=0.5]{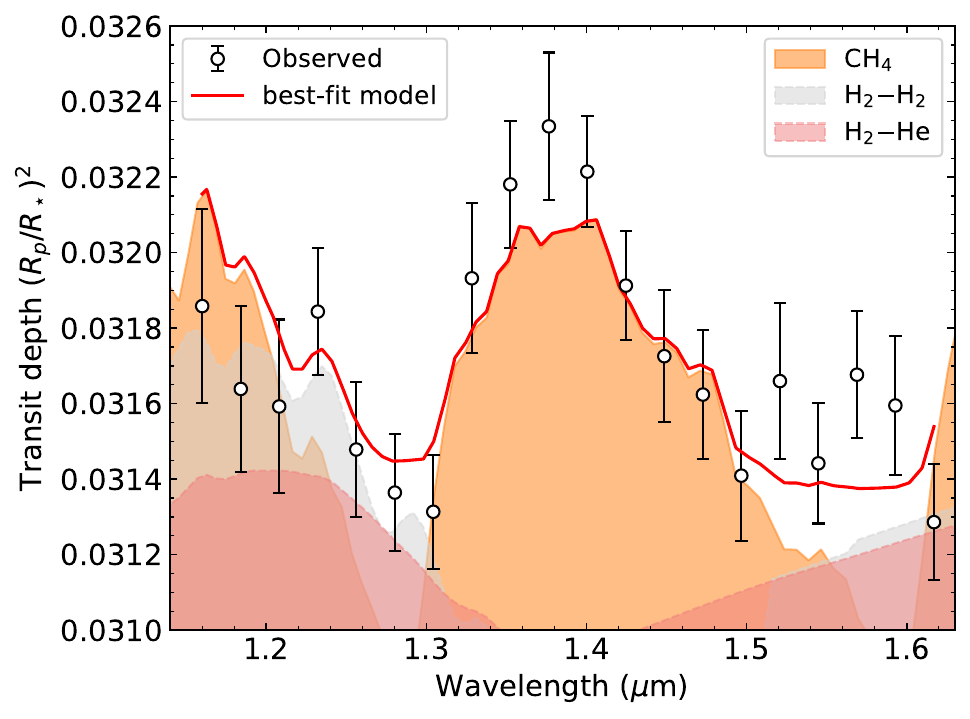}
\caption{Best-fit \ce{CH4}-only retrieval model of the HST transmission spectrum of TOI-6478 b. The wavelength-dependent opacity contributions from CH$_4$ and the gas continuum are shown. \ce{H2}--\ce{H2} and \ce{H2}--He continuum opacity are indicated by the gray and red shaded regions, respectively.}
\label{fig:prt_retrieval_ch4_only_bestfit_spectrum}
\end{figure}

\begin{deluxetable}{ll@{\hspace{0.25\columnwidth}}c}
\tablecaption{\\Full Atmospheric Model Results}
\label{tab:prt_full_model_results}
\setlength{\tabcolsep}{0pt}
\tablehead{
\multicolumn{1}{l}{Parameter} & & \colhead{Posterior Value}
}
\startdata
\addlinespace[3pt]
$R\mathrm{_p}$ (R$_\oplus$) & & $4.58^{+0.14}_{-0.07}$ \\
\addlinespace[3pt]
log$_{10}(g)$ & log$_{10}$(cm s$^{-2}$)  & $2.81^{+0.13}_{-0.13}$ \\
\addlinespace[3pt]
log$_{10}(P_\mathrm{ref}$)\,\,\,\,\, & log$_{10}$(bar) & $-1.83^{+2.05}_{-3}$ \\ 
\addlinespace[3pt]
$T$(K) & & $186^{+59.34}_{-48.54}$ \\
\addlinespace[3pt]
\ce{H2O} & log$_{10}$(mass frac) & $-2.58^{+0.58}_{-0.55}$ \\
\addlinespace[3pt]
\ce{CH4} & log$_{10}$(mass frac) & $-7.78^{+2.90}_{-2.62}$ \\
\addlinespace[3pt]
\ce{CO} & log$_{10}$(mass frac) & $-2.51^{+1.07}_{-5.44}$ \\
\addlinespace[3pt]
\ce{CO2} & log$_{10}$(mass frac) & $-6.71^{+3.69}_{-3.32}$ \\
\addlinespace[3pt]
\ce{NH3} & log$_{10}$(mass frac) & $-8.93^{+1.89}_{-1.87}$ \\
\addlinespace[3pt]
\ce{HCN} & log$_{10}$(mass frac) & $-6.73^{+1.84}_{-3.00}$ \\
\addlinespace[3pt]
$b$ & & $-7.40^{+1.36}_{-1.37}$ \\
\addlinespace[3pt]
\enddata
\tablenotetext{}{\textbf{Note:} Abundances in pRT are in units of mass fractions. One can convert between mass fractions and volume mixing ratios by using $n_i=(\mu/\mu_i)X_i$, Where $X_i$ is the mass fraction of species $i$, $\mu_i$ the molar mass of a species, and $\mu$ the atmosphere mean molar mass.} 
\end{deluxetable}

\begin{figure}
\centering
\includegraphics[scale=0.6]{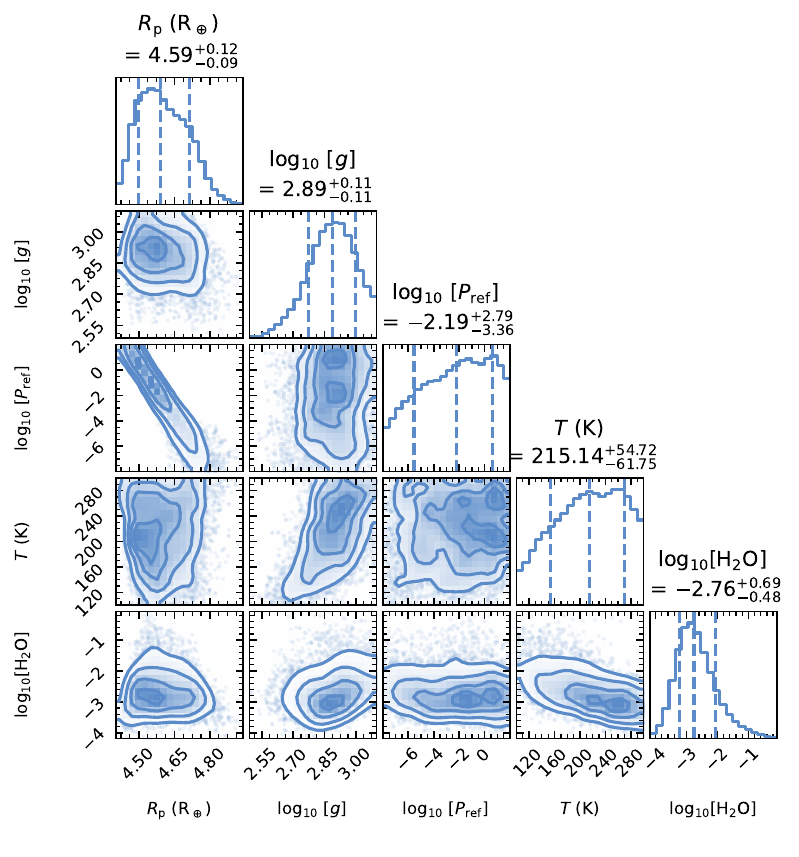}
\caption{The 2-D posterior distribution for the H$_2$O-only pRT retrieval model. From this fit we get constraints on the H$_2$O abundance providing us with statistical evidence that there is an observable atmosphere (see Section \ref{sec:discussion}). We also get a constraint on the $\log \rm g$ parameter allowing us to calculate a mass for TOI-6478 b.}
\label{fig:prt_h20_corner_plot}
\end{figure}

\begin{figure}
\centering
\includegraphics[scale=0.5]{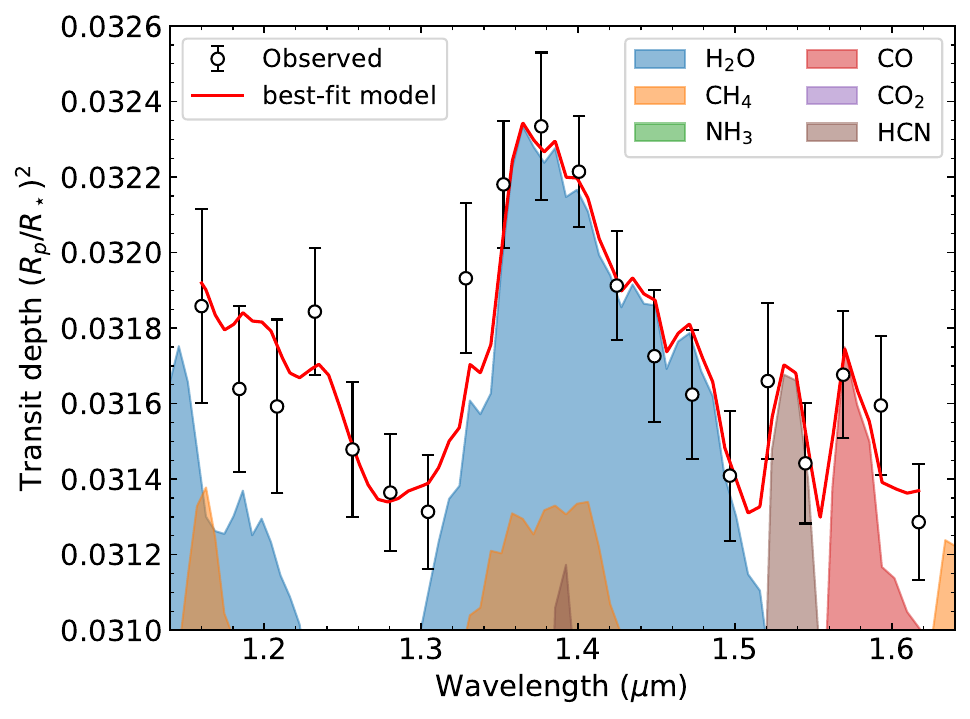}
\caption{Best-fit free-chemistry retrieval model of the HST transmission spectrum of TOI-6478 b, showing the opacity contributions of the full set of included molecular species as a function of wavelength. Gas continuum opacity from H$_2$--H$_2$ and H$_2$--He omitted from the plot.}
\label{fig:prt_retrieval_full_bestfit_spectrum}
\end{figure}

Table \ref{tab:prt_bayesian_evidence} shows the Bayesian evidence analysis results. We find that \ce{H2O} is the most favored molecule. We find little to no evidence for \ce{CH4} or \ce{NH3}. However, the \(\Delta \ln Z\) between models is small, indicating that the preference for H$_2$O is not statistically significant, and \ce{CH4} (or other molecules) cannot be ruled out. In Figure \ref{fig:prt_retrieval_h2o_only_bestfit_spectrum} we show the best-fit spectrum for the \ce{H2O}-only with a clear atmosphere and in Figure \ref{fig:prt_retrieval_ch4_only_bestfit_spectrum} we show the best-fit spectrum for \ce{CH4}-only with a clear atmosphere. Figure \ref{fig:prt_h20_corner_plot} shows the 2-D posterior distributions for the \ce{H2O}-only which is the most favored model. We also present the best-fit model spectrum and opacity contributions from the full set of key molecular species in  Figure \ref{fig:prt_retrieval_full_bestfit_spectrum}, the posterior median and 16th/84th percentile constraints in Table \ref{tab:prt_full_model_results}, and the 2-D posterior distributions in Figure \ref{fig:prt_full_corner_plot}.

\subsubsection{ExoTR free retrieval}\label{sec:ExoTR}

We analyzed the transmission spectrum of TOI‑6478~b using the Exoplanetary Transmission Retrieval framework \citep[\texttt{ExoTR},][]{damiano2024lhs}, which enables a statistical interpretation of observed exoplanet spectra. ExoTR generates forward transmission models and employs a nested sampling approach, implemented through \texttt{MultiNest} \citep{Feroz2009}, to efficiently explore a wide range of atmospheric configurations.

For TOI‑6478\,b, we treated the planet mass as a free parameter and adopted a uniform prior between 3 and 10 Earth masses in the retrievals, while imposing a uniform prior on the temperature between 100 and 500 K. In the reference model, we included H$_2$O, CH$_4$, CO, CO$_2$, NH$_3$, and HCN as fit molecules in an H$_2$-dominated atmosphere, as these molecules are broadly expected constituents of low-temperature sub-Neptune atmospheres \citep[e.g.,][]{yu2021identify, hu2021photochemistry, hu2021unveiling, tsai2021inferring}. Molecular opacities were computed line-by-line at a resolving power of 200,000, using HITEMP for CH$_4$ \citep{Hargreaves2020} and HITRAN2020 for all other species \citep{Gordon2022}. Additional opacity sources included H$_2$–H$_2$ collision-induced absorption and Rayleigh scattering, and we did not rescale the observational uncertainties in the retrieval analysis.

As summarized in Table~\ref{tab:exotr_scenarios}, we consider two cloud treatments: a ``gray cloud,'' in which the cloud opacity is wavelength-independent, and a ``physical cloud,'' in which cloud formation is governed by the condensation of water. In the latter case, the model converts the abundance of condensed water vapor into cloud mass -- reducing the corresponding amount from the above-cloud water vapor (i.e., the cold trap effect) -- and computes the corresponding wavelength-dependent opacity using Mie theory. Further details of the physical cloud implementation in \texttt{ExoTR} are provided by \cite{damiano2020exorel}. For each cloud scenario, we additionally toggle the TLSE \citep{rackham2018transit} on and off in the retrievals. The resulting four models yield comparable Bayesian evidences.

\begin{table}[h!]
\centering
\caption{Planetary scenarios explored with \texttt{ExoTR} for TOI‑6478~b. All reference models include six gases: H$_2$O, CH$_4$, NH$_3$, HCN, CO, and CO$_2$. The Bayes factor ($B$) for a gas is evaluated by comparing the evidence of the reference model with that of a model excluding the corresponding species. In scenarios with physical clouds, water condensation is assumed to drive cloud formation, thereby implicitly requiring the presence of water vapor in the atmosphere.}

\begin{tabular}{l l c c}
\hline
\textbf{Sce.} & \textbf{Config.} & \textbf{${\ln Z}$} & $B$ \\
\hline
\textbf{A} & \textbf{6 Gases+Gray Clds} & \textbf{135.1} & \\
 & -- w/o H$_2$O & 132.3 & 16 \\
 & -- w/o CH$_4$ & 135.2 & 0.9 \\
 & -- w/o H$_2$O and CH$_4$ & 114.6 & $8\times10^8$ \\
 & -- w/o CO & 134.6 & 1.6 \\
 \hline
\textbf{B} & \textbf{6 Gases+Gray Clds+TLSE} & \textbf{134.3} & \\
 & -- w/o H$_2$O & 133.8 & 1.6 \\
 & -- w/o CH$_4$ & 133.8 & 1.6 \\
 & -- w/o H$_2$O and CH$_4$ & 131.2 & 22 \\
 & -- w/o CO & 133.8 & 1.6 \\
 \hline
\textbf{C} & \textbf{6 Gases+Phys Clds} & \textbf{133.4} & \\
 & -- w/o CH$_4$ & 133.7 & 0.7 \\
 & -- w/o CO & 133.2 & 1.2 \\
 \hline
\textbf{D} & \textbf{6 Gases+Phys Clds+TLSE} & \textbf{134.0} & \\
 & -- w/o CH$_4$ & 133.0 & 2.7 \\
 & -- w/o CO & 133.7 & 1.3 \\
 \hline
\hline
\end{tabular}
\label{tab:exotr_scenarios}
\end{table}

Although no single scenario is strongly favored by the Bayesian evidence, the assumed configurations yield markedly different constraints on the atmospheric composition. In Scenario A, the retrieval shows a preference for H$_2$O, with no evidence for CH$_4$. When both H$_2$O and CH$_4$ are excluded, the resulting Bayes factor is very large, indicating a robust detection of at least one of these two species. Upon including the TLSE (Scenario B), the detection significance is reduced, and the relative preference between H$_2$O and CH$_4$ becomes comparable. The posterior distributions of their abundances exhibit a clear correlation, with mixing ratios for both species clustering around $10^{-5}$. In addition, H$_2$–H$_2$ collision-induced absorption contributes significantly to the observed spectrum in the best-fit model.

When adopting a physical cloud model driven by water condensation (Scenarios C and D), the preference for CH$_4$ is not strong, but the observed spectrum can be partially explained by CH$_4$ absorption, particularly when the TLSE effect is included (Scenario D). In the absence of CH$_4$, the retrieval favors a solution in which the deep atmospheric water mixing ratio is $\sim10^{-3}$, decreasing by $\sim2$ orders of magnitude near $\sim1$ bar to $\sim10^{-5}$ above the cloud deck. 

When CH$_4$ is included in the fit, however, the model permits arbitrarily strong water condensation (i.e., very low above-cloud water abundances), while the spectral modulation is primarily driven by CH$_4$ absorption. In this case, the inferred CH$_4$ mixing ratio is also $\sim10^{-5}$. Finally, across all scenarios, CO is invoked to account for features in the $1.55–1.60\ \mu$m range, although its detection significance remains weak. 

\subsection{Self-consistent forward models}\label{sec:forward_models}

We generate self-consistent models by applying the \texttt{EGP} framework \citep{marley1999reflected, fortney2008synthetic, morley2015thermal}, which calculates the temperature structures and abundances of atoms and molecules in substellar atmospheres. These models assume both radiative--convective equilibrium and chemical equilibrium, including the rainout of species that condense deeper within the atmosphere. We include irradiation from a host star using the PHOENIX model atmosphere grid \citep{husser2013new}. 

We include clouds in some models using the \texttt{eddysed} framework of \citet{ackerman2001precipitating}, which calculates the diffusive mixing of particles and gas upward and rate of particle settling downward, balancing the two mass fluxes to create an advective--diffusive balance. This approach allows us to calculate the location of the cloud and estimate the particle size in each layer. The species that condenses in this relatively cold planet is H$_2$O, which we assume condenses into ice. The free parameter $f_{\rm sed}$ describes the efficiency of sedimentation (lower sedimentation efficiency leads to more lofted clouds with smaller particle sizes). 

\begin{figure*}
\centering
\includegraphics[scale=0.36]{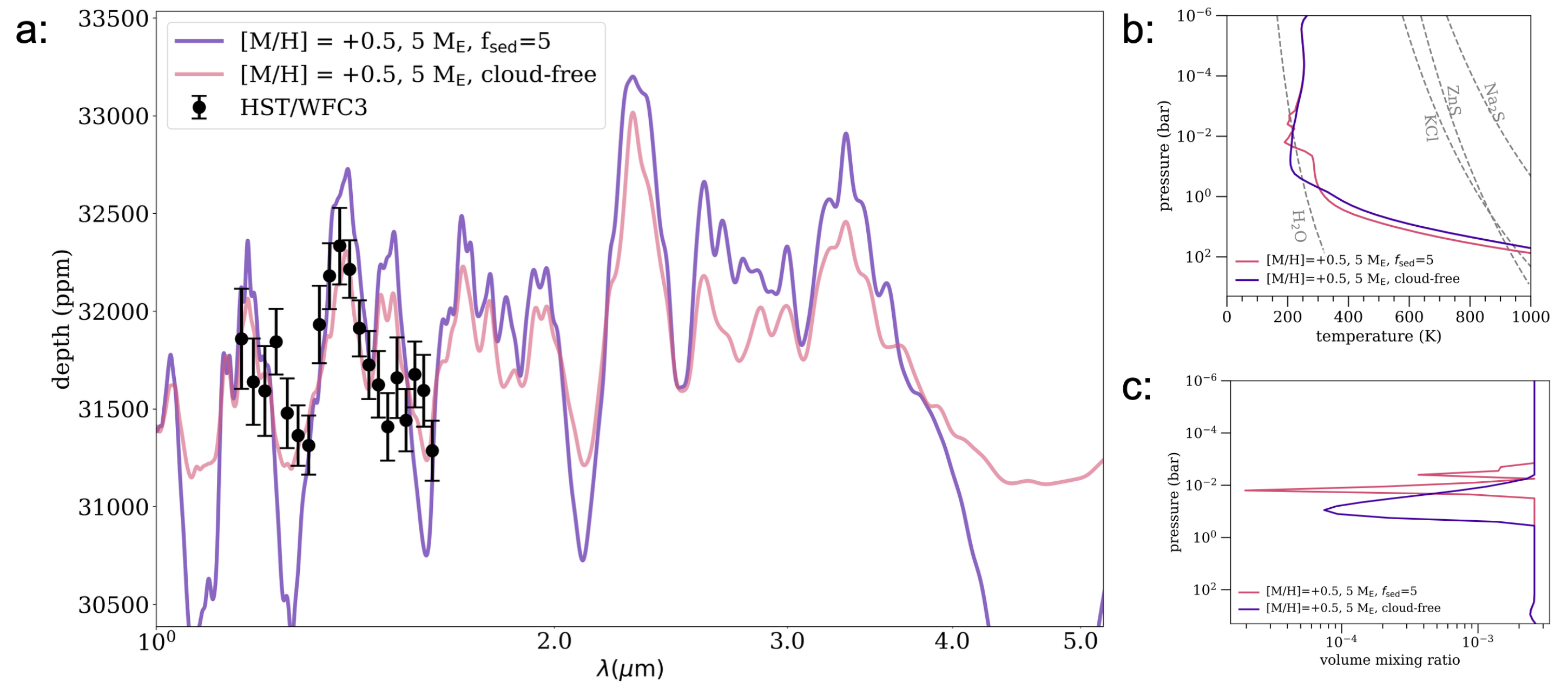}
\caption{\textbf{Panel a:} HST/WFC3 data plotted with the simulated transmission spectrum from the self-consistent forward model. The model parameters for these spectra are $5\rm \:M_{\oplus}$, $\rm M/H = +0.5$, and $\rm C/O = 1.0$, where the $f_{\rm sed}=5$ model is plotted in magenta and the cloud-free model plotted in purple. The points in black are the data from HST/WFC3 with 1 sigma error bars. \textbf{Panel b:} PT profiles from the self consistent forward models. Both lines use the same model parameters as in panel a. The gray dashed lines represent condensation curves for various condensables in the atmosphere. \textbf{Panel c:} Volume mixing ratio of H$_2$O from the self consistent forward models using the same models parameters as panels a and b.}
\label{fig:forward_model}
\end{figure*}

In Figure \ref{fig:forward_model}, we show the generated spectra with contribution from all gases and a cloud deck, as well as the HST/WFC3 data for reference. Our best-fit model shown in Figure \ref{fig:forward_model} has `thinner' clouds, with $f_{\rm sed}$=5. 
Although these models predict water-ice formation and spectra dominated by CH$_4$ in the HST bandpass, they also retain a contribution from water vapor at high altitudes. This modeling therefore indicates that, while condensation is expected to suppress water gas absorption at these temperatures, it may not remove it entirely. This leaves open the possibility that H$_2$O contributes to, or produces, the observed feature even in this cold atmosphere.

To explain why H$_2$O both forms a cloud \emph{and} appears as a gas-phase absorber, we consider the pressure-temperature (PT) structure of the model atmosphere which we show in Figure \ref{fig:forward_model}. We can calculate where the vapor pressure of H$_2$O exceeds the saturation vapor pressure of water, which it does around $10^{-1}$ bar in a cloud-free model and $10^{-2}$ bar in a cloudy model, where both the albedo of the cloud and infrared opacity of the cloud change the structure of the atmosphere in a self-consistent model radiative--convective model including cloud opacity. However, the upper atmosphere, heated by the host star, is warmer than the $10^{-1}$--$10^{-2}$ bar level where water condensation occurs. Above a few mbar, the vapor pressure of water is \emph{less} than the saturation vapor pressure. 

In Figure \ref{fig:forward_model} we show the volume mixing ratio of H$_2$O from the self-consistent forward models. While this increase in water abundance in the upper atmosphere occurs in the model, this is not the behavior of water above the convective weather layer on Earth. Earth's water is `cold-trapped' by condensation at the coldest point in the atmosphere, with water clouds forming there and a relatively dry stratosphere above \citep{wordsworth2013water, madhusudhan2025exploring}. Whether this mechanism occurs in the atmosphere of TOI-6478 b is not known, but can be tested by future observations that disentangle water and methane absorption. We note that condensables can also be cold-trapped on the night-side of tidally-locked planets, further complicating this picture \citep{parmentier20133d, leconte20133d}. 

\section{Discussion}\label{sec:discussion}

Sub-Neptune exoplanets have been studied extensively with space-based telescopes such as HST and JWST. For the most part, their atmospheres have eluded compositional characterization, due to the formation of clouds and hazes that reduce atmospheric clarity. Among the current population, only a few sub-Neptunes have detectable atmospheres with identified carbon- and oxygen-bearing molecules \citep{hu2025water, benneke2024jwst, beatty2024sulfur}. TOI-6478 b is the latest sub-Neptune found to have an observable atmosphere: its HST/WFC3 transmission spectrum shows a clear absorption feature, and our free retrievals provide statistical evidence that this is due to absorption from either H$_2$O or CH$_4$. TOI-6478 b is also the coldest sub-Neptune ($\rm T_{eq} \sim200\:K$) with a detected atmosphere, and so offers an opportunity to study a unique region of parameter space. As the sample of temperate planets with detected atmospheres grows, observations like these will improve our understanding of the physics of cold sub-Neptune atmospheres.

As mentioned, a recent analysis by \cite{brande2024clouds} studied the atmospheric clarity of sub-Neptunes by measuring the spectral feature amplitudes of their WFC3 transmission spectra as a function of equilibrium temperature, and comparing these to atmospheric models with several sedimentation efficiencies and haze particle sizes. They found that the planets with low spectral amplitudes (flat transmission spectra) fall in the range of $\mathrm{500 \: K <T_{eq} < 700 \:K}$, and that these planets were well matched by atmospheric models that had high metallicities and low sedimentation efficiencies. \cite{brande2024clouds} also found that planets on both the cooler ($\mathrm{T_{eq} < 500\:K}$) and hotter ($\mathrm{T_{eq}>700\:K}$) sides of this temperature range had significantly less clouds in their atmospheres, and correspondingly larger absorption features in their transmission spectra. The simplest explanation for this ``quadratic'' trend in cloudiness was caused by the warmer planets being too hot for significant clouds to form \citep{helling2021cloud}, the intermediate planets having low sedimentation efficiencies and extended, visible, cloud decks, and the cooler planets having high sedimentation efficiencies and cloud decks that have sunk below the visible photosphere.

This physical interpretation mirrors the brown dwarf cloud sequence. As the effective temperature of a brown dwarf falls below $\mathrm{T_{eff} \approx 1300\: K}$, changes in their spectra suggest that clouds disappear below the visible photosphere, marking the L/T transition \citep{buenzli2014brown}. The proposed mechanism for this is cloud thinning, in which increasing cloud particle sizes lead to a rapid rainout that thins and eventually removes the visible clouds \citep{tsuji2003transition, knapp2004near}. 

\cite{brande2024clouds} noted that planets with $\mathrm{T_{eq}<500\: K}$ remained poorly characterized, and that observing more of these temperate planets should be a priority to better understand the trend of atmospheric clarity versus equilibrium temperature. TOI-6478 b is one such planet, and the strong absorption feature in its transmission spectrum is consistent with the high sedimentation efficiencies and lack of visible clouds expected at these temperatures. Its low temperature, however, also complicates the identification of the absorber. The best-fit model from our free retrievals finds a temperature of $\mathrm{T = 186^{+59.34}_{-48.54}\:K}$, and H$_2$O at this temperature should condense or be cold-trapped below the observable photosphere. Our free retrievals nevertheless favor H$_2$O, though not at a statistically significant level, while our self-consistent forward models suggest that CH$_4$ could dominate the spectrum. This raises the question of whether the absorption in the HST/WFC3 bandpass is due to H$_2$O or CH$_4$, and if it is H$_2$O, how does it persist in the gas phase at these low temperatures? 

If the absorber is H$_2$O, then either its condensation is limited -- either by a low cloud nucleation efficiency or by an inefficient cold trap -- or the planet is internally heated and gas-phase H$_2$O is being lofted into the cool upper atmosphere from the hotter interior \citep[e.g.,][]{Welbanks2024}. The existing radial velocity (RV) observations of the TOI-6478 system do not yet significantly constrain the orbit of TOI-6478 b, so its eccentricity is unknown. If future RV measurements reveal a significant orbital eccentricity, then a confirmation of a significant H$_2$O abundance would likely indicate that TOI-6478 b is internally heated by tidal forces \citep{Welbanks2024}. The presence of H$_2$O in TOI-6478 b's atmosphere would therefore be suggestive of tidally driven radius-inflation, which would also naturally explain the planet's relatively low density. 

\begin{figure}
\centering
\includegraphics[scale=0.38]{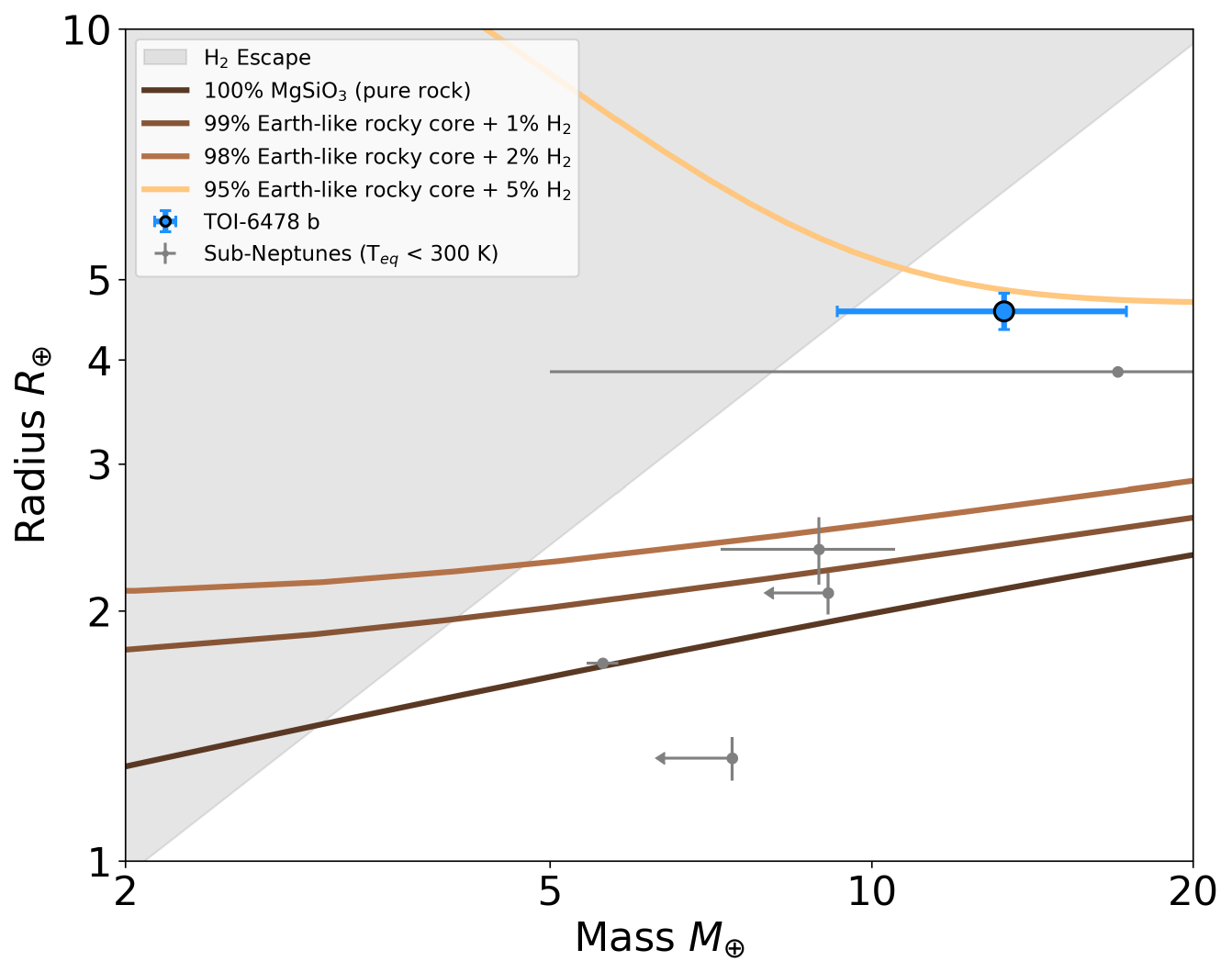}
\caption{We plot mass versus radius curves with Earth-like rocky cores and increasing H$_2$ envelopes from \cite{zeng2019growth}. The gray shaded region is the H$_2$ escape region calculated for TOI-6478 b using Equation 22 from \cite{aguichine2021} assuming $(v_{\mathrm{esc}}/v_{\mathrm{th}})^2 \approx100$. The gray points are the current population of temperate sub-Neptunes with $\rm T_{eq}<300 \: K$ and the light blue point is TOI-6478 b's mass and radius constraints from the free chemistry retrievals.}
\label{fig:Mass_v_Radius}
\end{figure}

\begin{figure*}
\centering
\includegraphics[scale=0.65]{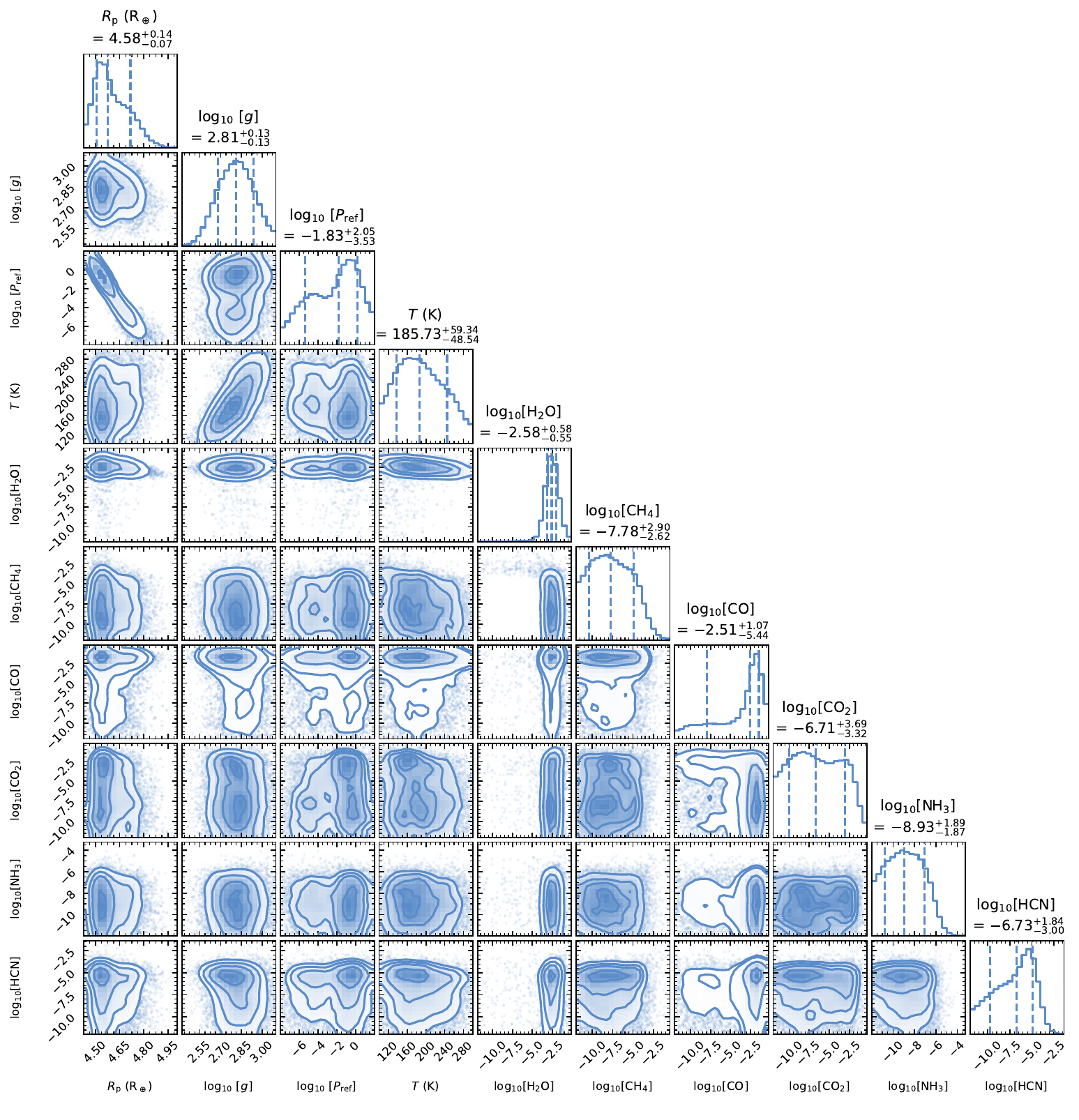}
\caption{The 2-D posteriors for the pRT retrieval model with all absorbers included. Similar to Figure \ref{fig:prt_h20_corner_plot}, we get constraints on an H$_2$O abundance and the $\log \rm g$ parameter. The full retrieval model provides statistical evidence for an H$_2$O dominated atmosphere over a CH$_4$ dominated atmosphere. Additionally, we get some evidence for contribution from CO however, the long tail at the end of the posterior distribution suggests this abundance is unconstrained. With the $\log \rm (g)$ constraint we calculate a mass of $\rm M = 13.3\pm4.02 \: M_{\oplus}$ which places TOI-6478 b in a unique parameter space relative to other cold ($\rm T_{eq}<300\:K$) sub-Neptunes (see Section \ref{sec:discussion}).}
\label{fig:prt_full_corner_plot}
\end{figure*}

Further, from our free chemistry retrievals we were able to get a measurement for the mass of TOI-6478 b by fitting for the $\log \rm (g)$ parameter. After propagating error, we a get a mass of $\rm M = 13.3\pm4.02 \: M_{\oplus}$. In Figure \ref{fig:Mass_v_Radius} we plot various mass vs radius curves with Earth-like rocky cores and increasing H$_2$ envelopes from \cite{zeng2019growth} as well as the H$_2$ escape region calculated using Equation 22 from \cite{aguichine2021}. From this we conclude that TOI-6478 b's mass and radius measurements are indicative of a planet with a massive hydrogen envelope puffing up the atmosphere. For reference, we plot the only other temperate sub-Neptunes ($\rm T_{eq}<300\: K$) that have mass and radius measurements to demonstrate that TOI-6478 b exists outside of the range we would expect for a sub-Neptune of this mass. However, given the lack of temperate sub-Neptune's with mass and radius measurements in the exoplanet catalog, it is yet to be seen if TOI-6478 b exists in a unique parameter space. 

Finally, we note that the disagreement between our retrievals and forward models demonstrates that the HST/WFC3 bandpass alone cannot distinguish between these two scenarios, as has been recognized in other HST/WFC3 transit observations \citep{ehrenreich2014near, murphy2025hst, benneke2019sub, blain20211d}. Determining the composition of TOI-6478 b's atmosphere with JWST would provide new insight into the cloud sequence of sub-Neptunes, and would allow us to trace physical processes, such as tidal heating and vertical mixing, that would otherwise be unobservable. It would also provide the elemental abundance ratios (O/H, C/H) that can be used to infer the formation and migration pathways of this sub-Neptune.

\section{Conclusions}

We have presented the near-IR transmission spectrum of the cold ($\mathrm{T_{eq}\approx 200\:K}$) sub-Neptune TOI-6478\,b, observed in two visits with the HST/WFC3 G141 grism. We fit both visits simultaneously to produce the transmission spectrum, and performed three atmospheric analyses to determine the composition of TOI-6478 b's atmosphere: free retrievals, self-consistent forward models, and tests of the sensitivity of our results to the treatment of TLSE and clouds. The spectrum shows a clear absorption feature within the WFC3 bandpass of 1.1 to 1.7 $\mu$m, and all three analyses significantly detect an absorption feature in the data. Due to the limited wavelength range of the WFC3 bandpass, our analysis is only able to narrow the responsible absorber down to H$_2$O and CH$_4$, which we find are not distinguishable using these data.

Our free retrievals find that H$_2$O is the most favored molecule, with nominally insignificant abundances of CH$_4$ or NH$_3$. However, the model preference for H$_2$O is not statistically significant, given that the \(\Delta \ln Z\) between the free retrievals in which the H$_2$O abundance is forced to zero and which instead find significant CH$_4$ is small. This is mainly a result of the CH$_4$ and H$_2$O absorption features overlaying each other near 1.4 $\mu$m. Our self-consistent forward models show that TOI-6478 b is in the temperature regime where should water condense, which would remove it from the upper atmosphere. The forward models instead prefer CH$_4$, with limited H$_2$O and NH$_3$ abundances. Our tests of the TLSE and various cloud treatments show that the inferred detection significance is mildly sensitive to the inclusion of stellar heterogeneity, but even with TLSE and clouds included the models are unable to disentangle the contributions of H$_2$O and CH$_4$.

If H$_2$O is responsible for the observed spectral feature, then its condensation must be inhibited -- either by limited cloud nucleation efficiency or by an inefficient cold trap -- or the planet must be internally heated, for example by tides if its orbit proves to be eccentric. Alternatively, if H$_2$O is strongly depleted in the upper atmosphere, the feature can be explained by CH$_4$. This requires only a relatively small, naturally occurring, CH$_4$ abundance. Lifting this degeneracy requires JWST observations that cover a wider wavelength range than the WFC3 bandpass. TOI-6478 b is the coldest sub-Neptune with an atmosphere detected in transmission, and these results identify it as an excellent testbed for several atmospheric scenarios. 

\section*{Acknowledgments}

This research is based on observations made with the NASA/ESA Hubble Space Telescope obtained from the Space Telescope Science Institute, which is operated by the Association of Universities for Research in Astronomy, Inc., under NASA contract NAS 5–26555. We would like to thank the HST staff for helping us to move quickly to observe this newly-identified target.

AT, JP and IJMC acknowledge funding support from STScI as part of this observing program.

MGS acknowledges support from the UK Science and Technology Facilities Council (STFC) and from a local studentship delivered by the College of Engineering and Physical Sciences of the University of Birmingham.
KAK gratefully acknowledges support from the DLR via project P.S.ASTR1508.

This work has made use of NASA's Astrophysics Data System. This research has made use of the NASA Exoplanet Archive, which is operated by the California Institute of Technology, under contract with the National Aeronautics and Space Administration under the Exoplanet Exploration Program.

\clearpage
\bibliography{bib}

@MISC{debesetal24-1,
       author = {{Debes}, John and {Sankrit}, Ravi and {Fischer}, Travis and {Frazer}, Elaine and {Hirschauer}, Alec and {Rowlands}, Kate and {Burger}, Matthew and {Swaters}, Robert and {Jedrzejewski}, Robert and {Gomez}, Sierrra and {Dos Santos}, Leonardo and {Hernandez}, Svea and {Miller}, Lauren and {Payne}, Anna and {Rafelski}, Marc and {Wevers}, Thomas and {Anderson}, Sara and {Bair}, Tom and {Bello}, Kathryn and {Carlberg}, Joleen and {Charlow}, Brian and {Cortese}, Andrew and {Dencheva}, Nadia and {Ellis}, Tracy and {Falk}, Ben and {Fleming}, Scott and {Forshay}, Peter and {Gilani}, Syed and {Hall}, Patty and {Kimball}, Tim and {Kelley}, Talya and {Kidwell}, Richard and {Kotler}, Jenn and {Kovacs}, Aiden and {James}, Bethan and {Rahmani}, Christopher and {Rodriguez}, David and {Roman-Duval}, Julia and {Soderblom}, David and {Sherbert}, Lisa and {Welty}, Dan and {Wolfe}, David},
        title = "{The Hubble Advanced Spectral Product (HASP) Program}",
 howpublished = {Instrument Science Report COS 2024-01, 31 pages},
         year = 2024,
        month = jan,
        pages = {1},
       adsurl = {https://ui.adsabs.harvard.edu/abs/2024cos..rept....1D}
}

@ARTICLE{woodgateetal98-1,
       author = {{Woodgate}, B.~E. and {Kimble}, R.~A. and {Bowers}, C.~W. and
         {Kraemer}, S. and {Kaiser}, M.~E. and {Danks}, A.~C. and
         {Grady}, J.~F. and {Loiacono}, J.~J. and {Brumfield}, M. and
         {Feinberg}, L. and {Gull}, T.~R. and {Heap}, S.~R. and {Maran}, S.~P. and
         {Lindler}, D. and {Hood}, D. and {Meyer}, W. and {Vanhouten}, C. and
         {Argabright}, V. and {Franka}, S. and {Bybee}, R. and {Dorn}, D. and
         {Bottema}, M. and {Woodruff}, R. and {Michika}, D. and {Sullivan}, J. and
         {Hetlinger}, J. and {Ludtke}, C. and {Stocker}, R. and {Delamere}, A. and
         {Rose}, D. and {Becker}, I. and {Garner}, H. and {Timothy}, J.~G. and
         {Blouke}, M. and {Joseph}, C.~L. and {Hartig}, G. and {Green}, R.~F. and
         {Jenkins}, E.~B. and {Linsky}, J.~L. and {Hutchings}, J.~B. and
         {Moos}, H.~W. and {Boggess}, A. and {Roesler}, F. and {Weistrop}, D.},
        title = "{The Space Telescope Imaging Spectrograph Design}",
      journal = {\pasp},
         year = "1998",
        month = "Oct",
       volume = {110},
       number = {752},
        pages = {1183-1204},
          doi = {10.1086/316243},
       adsurl = {https://ui.adsabs.harvard.edu/abs/1998PASP..110.1183W}
}

@article{damiano2020exorel,
  title={Exorel: a Bayesian inverse retrieval framework for exoplanetary reflected light spectra},
  author={Damiano, Mario and Hu, Renyu},
  journal={The Astronomical Journal},
  volume={159},
  number={4},
  pages={175},
  year={2020},
  publisher={The American Astronomical Society}
}

@article{tsai2021inferring,
  title={Inferring Shallow Surfaces on Sub-Neptune Exoplanets with JWST},
  author={Tsai, Shang-Min and Innes, Hamish and Lichtenberg, Tim and Taylor, Jake and Malik, Matej and Chubb, Katy and Pierrehumbert, Raymond},
  journal={The Astrophysical Journal Letters},
  volume={922},
  number={2},
  pages={L27},
  year={2021},
  publisher={IOP Publishing}
}

@article{hu2021photochemistry,
  title={Photochemistry and spectral characterization of temperate and gas-rich exoplanets},
  author={Hu, Renyu},
  journal={The Astrophysical Journal},
  volume={921},
  number={1},
  pages={27},
  year={2021},
  publisher={IOP Publishing}
}

@article{hu2021unveiling,
  title={Unveiling shrouded oceans on temperate sub-Neptunes via transit signatures of solubility equilibria versus gas thermochemistry},
  author={Hu, Renyu and Damiano, Mario and Scheucher, Markus and Kite, Edwin and Seager, Sara and Rauer, Heike},
  journal={The Astrophysical Journal Letters},
  volume={921},
  number={1},
  pages={L8},
  year={2021},
  publisher={IOP Publishing}
}

@article{yu2021identify,
  title={How to identify exoplanet surfaces using atmospheric trace species in hydrogen-dominated atmospheres},
  author={Yu, Xinting and Moses, Julianne I and Fortney, Jonathan J and Zhang, Xi},
  journal={The Astrophysical Journal},
  volume={914},
  number={1},
  pages={38},
  year={2021},
  publisher={IOP Publishing}
}

@ARTICLE{Gordon2022,
       author = {{Gordon}, I.~E. and {Rothman}, L.~S. and {Hargreaves}, R.~J. and {Hashemi}, R. and {Karlovets}, E.~V. and {Skinner}, F.~M. and {Conway}, E.~K. and {Hill}, C. and {Kochanov}, R.~V. and {Tan}, Y. and {Wcis{\l}o}, P. and {Finenko}, A.~A. and {Nelson}, K. and {Bernath}, P.~F. and {Birk}, M. and {Boudon}, V. and {Campargue}, A. and {Chance}, K.~V. and {Coustenis}, A. and {Drouin}, B.~J. and {Flaud}, J. -M. and {Gamache}, R.~R. and {Hodges}, J.~T. and {Jacquemart}, D. and {Mlawer}, E.~J. and {Nikitin}, A.~V. and {Perevalov}, V.~I. and {Rotger}, M. and {Tennyson}, J. and {Toon}, G.~C. and {Tran}, H. and {Tyuterev}, V.~G. and {Adkins}, E.~M. and {Baker}, A. and {Barbe}, A. and {Can{\`e}}, E. and {Cs{\'a}sz{\'a}r}, A.~G. and {Dudaryonok}, A. and {Egorov}, O. and {Fleisher}, A.~J. and {Fleurbaey}, H. and {Foltynowicz}, A. and {Furtenbacher}, T. and {Harrison}, J.~J. and {Hartmann}, J. -M. and {Horneman}, V. -M. and {Huang}, X. and {Karman}, T. and {Karns}, J. and {Kassi}, S. and {Kleiner}, I. and {Kofman}, V. and {Kwabia-Tchana}, F. and {Lavrentieva}, N.~N. and {Lee}, T.~J. and {Long}, D.~A. and {Lukashevskaya}, A.~A. and {Lyulin}, O.~M. and {Makhnev}, V. Yu. and {Matt}, W. and {Massie}, S.~T. and {Melosso}, M. and {Mikhailenko}, S.~N. and {Mondelain}, D. and {M{\"u}ller}, H.~S.~P. and {Naumenko}, O.~V. and {Perrin}, A. and {Polyansky}, O.~L. and {Raddaoui}, E. and {Raston}, P.~L. and {Reed}, Z.~D. and {Rey}, M. and {Richard}, C. and {T{\'o}bi{\'a}s}, R. and {Sadiek}, I. and {Schwenke}, D.~W. and {Starikova}, E. and {Sung}, K. and {Tamassia}, F. and {Tashkun}, S.~A. and {Vander Auwera}, J. and {Vasilenko}, I.~A. and {Vigasin}, A.~A. and {Villanueva}, G.~L. and {Vispoel}, B. and {Wagner}, G. and {Yachmenev}, A. and {Yurchenko}, S.~N.},
        title = "{The HITRAN2020 molecular spectroscopic database}",
      journal = {\jqsrt},
         year = 2022,
        month = jan,
       volume = {277},
          eid = {107949},
        pages = {107949},
          doi = {10.1016/j.jqsrt.2021.107949},
       adsurl = {https://ui.adsabs.harvard.edu/abs/2022JQSRT.27707949G}
}

@ARTICLE{Hargreaves2020,
       author = {{Hargreaves}, Robert J. and {Gordon}, Iouli E. and {Rey}, Michael and {Nikitin}, Andrei V. and {Tyuterev}, Vladimir G. and {Kochanov}, Roman V. and {Rothman}, Laurence S.},
        title = "{An Accurate, Extensive, and Practical Line List of Methane for the HITEMP Database}",
      journal = {\apjs},
         year = 2020,
        month = apr,
       volume = {247},
       number = {2},
          eid = {55},
        pages = {55},
          doi = {10.3847/1538-4365/ab7a1a},
archivePrefix = {arXiv},
       eprint = {2001.05037},
 primaryClass = {astro-ph.EP},
       adsurl = {https://ui.adsabs.harvard.edu/abs/2020ApJS..247...55H}
}

@article{rackham2018transit,
  title={The transit light source effect: false spectral features and incorrect densities for M-dwarf transiting planets},
  author={Rackham, Benjamin V and Apai, D{\'a}niel and Giampapa, Mark S},
  journal={The Astrophysical Journal},
  volume={853},
  number={2},
  pages={122},
  year={2018},
  publisher={The American Astronomical Society}
}

@article{Feroz2009,
	author = {Feroz, F. and Hobson, M. P. and Bridges, M.},
	doi = {10.1111/j.1365-2966.2009.14548.x},
	eprint = {https://academic.oup.com/mnras/article-pdf/398/4/1601/3039078/mnras0398-1601.pdf},
	issn = {0035-8711},
	journal = {Monthly Notices of the Royal Astronomical Society},
	month = {09},
	number = {4},
	pages = {1601-1614},
	title = {{MultiNest: an efficient and robust Bayesian inference tool for cosmology and particle physics}},
	url = {https://doi.org/10.1111/j.1365-2966.2009.14548.x},
	volume = {398},
	year = {2009}
}

@article{damiano2024lhs,
  title={LHS 1140 b is a potentially habitable water world},
  author={Damiano, Mario and Bello-Arufe, Aaron and Yang, Jeehyun and Hu, Renyu},
  journal={The Astrophysical Journal Letters},
  volume={968},
  number={2},
  pages={L22},
  year={2024},
  publisher={The American Astronomical Society}
}

@article{horne1986optimal,
  title={An optimal extraction algorithm for CCD spectroscopy.},
  author={Horne, Keith},
  journal={Publications of the Astronomical Society of the Pacific},
  volume={98},
  number={604},
  pages={609--617},
  year={1986},
  publisher={The Astronomical Society of the Pacific}
}

@article{allard2011stellar,
  title={Stellar to Substellar Model Atmospheres},
  author={Allard, France and Homeier, Derek and Freytag, Bernd},
  journal={Proceedings of the International Astronomical Union},
  volume={7},
  number={S282},
  pages={235--242},
  year={2011},
  publisher={Cambridge University Press}
}

@article{scott2025toi,
  title={TOI-6478 b: a cold underdense Neptune transiting a fully convective M dwarf from the thick disc},
  author={Scott, Madison G and Triaud, Amaury HMJ and Barkaoui, Khalid and Sebastian, Daniel and Burgasser, Adam J and Collins, Karen A and Dransfield, Georgina and Hellier, Coel and Howell, Steve B and Piette, Anjali AA and others},
  journal={Monthly Notices of the Royal Astronomical Society},
  volume={540},
  number={2},
  pages={1909--1927},
  year={2025},
  publisher={Oxford University Press}
}

@article{Grant2024,
  title = {ExoTiC-LD: thirty seconds to stellar limb-darkening coefficients},
  author = {David Grant and Hannah R. Wakeford},
  journal = {Journal of Open Source Software},
  publisher = {The Open Journal},
  year = {2024},
  volume = {9},
  number = {100},
  pages = {6816},
  doi = {10.21105/joss.06816},
  url = {https://doi.org/10.21105/joss.06816}
}

@article{kostogryz2023mps,
  title={MPS-ATLAS library of stellar model atmospheres and spectra},
  author={Kostogryz, N and Shapiro, AI and Witzke, V and Grant, D and Wakeford, HR and Stevenson, KB and Solanki, SK and Gizon, L},
  journal={Research Notes of the AAS},
  volume={7},
  number={3},
  pages={39},
  year={2023},
  publisher={The American Astronomical Society}
}

@article{kreidberg2015batman,
  title={batman: basic transit model calculation in Python},
  author={Kreidberg, Laura},
  journal={Publications of the Astronomical Society of the Pacific},
  volume={127},
  number={957},
  pages={1161--1165},
  year={2015},
  publisher={University of Chicago Press}
}

@article{Blain:2024prt,
    author = {Blain, Doriann and Molli\`ere, Paul and Nasedkin, Evert}, 
    title = {SpectralModel: a high-resolution framework for petitRADTRANS 3}, 
    journal = {The Journal of Open Source Software}, 
    volume = {9}, 
    number = {102}, 
    pages = {7028}, 
    year = {2024},
}

@article{Molliere:2019prt,
    author = {Molli\`ere, P. and Wardenier, J.~P. and van Boekel, R. and Henning, Th. and Molaverdikhani, K. and Snellen, I.~A.~G.}, 
    title = {petitRADTRANS. A Python radiative transfer package for exoplanet characterization and retrieval}, 
    journal = {\aap}, 
    volume = {627}, 
    pages = {A67}, 
    year = {2019},
}

@article{Nasedkin:2024prt,
    author = {Nasedkin, Evert and Molli\`ere, Paul and Blain, Doriann}, 
    title = {Atmospheric Retrievals with petitRADTRANS}, 
    journal = {The Journal of Open Source Software}, 
    volume = {9}, 
    number = {96}, 
    pages = {5875}, 
    year = {2024},
}

@ARTICLE{Feroz:2009multinest,
    author = {{Feroz}, F. and {Hobson}, M.~P. and {Bridges}, M.},
    title = "{MULTINEST: an efficient and robust Bayesian inference tool for cosmology and particle physics}",
    journal = {\mnras},
     year = 2009,
    month = oct,
    volume = {398},
    number = {4},
    pages = {1601-1614},
      doi = {10.1111/j.1365-2966.2009.14548.x},
    archivePrefix = {arXiv},
    eprint = {0809.3437},
    primaryClass = {astro-ph},
    adsurl = {https://ui.adsabs.harvard.edu/abs/2009MNRAS.398.1601F}
}

@ARTICLE{Buchner:2014_PyMultinest,
    author = {{Buchner}, J. and {Georgakakis}, A. and {Nandra}, K. and {Hsu}, L. and {Rangel}, C. and {Brightman}, M. and {Merloni}, A. and {Salvato}, M. and {Donley}, J. and {Kocevski}, D.},
    title = "{X-ray spectral modelling of the AGN obscuring region in the CDFS: Bayesian model selection and catalogue}",
    journal = {\aap},
     year = 2014,
    month = apr,
    volume = {564},
      eid = {A125},
    pages = {A125},
      doi = {10.1051/0004-6361/201322971},
    archivePrefix = {arXiv},
    eprint = {1402.0004},
    primaryClass = {astro-ph.HE},
    adsurl = {https://ui.adsabs.harvard.edu/abs/2014A&A...564A.125B}
}

@Article{Line:2015,
    author = {Line, Michael R. and Teske, Johanna and Burningham, Ben and Fortney, Jonathan J. and Marley, Mark S.}, 
    title = {Uniform Atmospheric Retrieval Analysis of Ultracool Dwarfs. I. Characterizing Benchmarks, Gl 570D and HD 3651B}, 
    journal = {\apj}, 
    volume = {807}, 
    number = {2}, 
    pages = {183}, 
    year = {2015},
}

@ARTICLE{Molliere:2025,
       author = {{Molli{\`e}re}, P. and {K{\"u}hnle}, H. and {Matthews}, E.~C. and {Henning}, Th. and {Min}, M. and {Patapis}, P. and {Lagage}, P.-O. and {Waters}, L.~B.~F.~M. and {G{\"u}del}, M. and {J{\"a}ger}, C. and {Zhang}, Z. and {Decin}, L. and {Biller}, B.~A. and {Absil}, O. and {Argyriou}, I. and {Barrado}, D. and {Cossou}, C. and {Glasse}, A. and {Olofsson}, G. and {Pye}, J.~P. and {Rouan}, D. and {Samland}, M. and {Scheithauer}, S. and {Tremblin}, P. and {Whiteford}, N. and {van Dishoeck}, E.~F. and {{\"O}stlin}, G. and {Ray}, T.},
        title = "{Evidence for SiO cloud nucleation in the rogue planet PSO J318}",
      journal = {\aap},
         year = 2025,
        month = nov,
       volume = {703},
          eid = {A79},
        pages = {A79},
          doi = {10.1051/0004-6361/202555732},
archivePrefix = {arXiv},
       eprint = {2507.18691},
 primaryClass = {astro-ph.EP},
       adsurl = {https://ui.adsabs.harvard.edu/abs/2025A&A...703A..79M}
}

@ARTICLE{Polyansky:2018,
       author = {{Polyansky}, Oleg L. and {Kyuberis}, Aleksandra A. and {Zobov}, Nikolai F. and {Tennyson}, Jonathan and {Yurchenko}, Sergei N. and {Lodi}, Lorenzo},
        title = "{ExoMol molecular line lists XXX: a complete high-accuracy line list for water}",
      journal = {\mnras},
         year = 2018,
        month = oct,
       volume = {480},
       number = {2},
        pages = {2597-2608},
          doi = {10.1093/mnras/sty1877},
archivePrefix = {arXiv},
       eprint = {1807.04529},
 primaryClass = {astro-ph.EP},
       adsurl = {https://ui.adsabs.harvard.edu/abs/2018MNRAS.480.2597P}
}

@ARTICLE{Yurchenko:2024,
       author = {{Yurchenko}, Sergei N. and {Owens}, Alec and {Kefala}, Kyriaki and {Tennyson}, Jonathan},
        title = "{ExoMol line lists - LVII. High accuracy ro-vibrational line list for methane (CH$_{4}$)}",
      journal = {\mnras},
         year = 2024,
        month = feb,
       volume = {528},
       number = {2},
        pages = {3719-3729},
          doi = {10.1093/mnras/stae148},
       adsurl = {https://ui.adsabs.harvard.edu/abs/2024MNRAS.528.3719Y}
}

@ARTICLE{Yurchenko:2020,
       author = {{Yurchenko}, S.~N. and {Mellor}, Thomas M. and {Freedman}, Richard S. and {Tennyson}, J.},
        title = "{ExoMol line lists - XXXIX. Ro-vibrational molecular line list for CO$_{2}$}",
      journal = {\mnras},
         year = 2020,
        month = aug,
       volume = {496},
       number = {4},
        pages = {5282-5291},
          doi = {10.1093/mnras/staa1874},
archivePrefix = {arXiv},
       eprint = {2007.02122},
 primaryClass = {astro-ph.EP},
       adsurl = {https://ui.adsabs.harvard.edu/abs/2020MNRAS.496.5282Y}
}

@ARTICLE{Rothman:2010,
       author = {{Rothman}, L.~S. and {Gordon}, I.~E. and {Barber}, R.~J. and {Dothe}, H. and {Gamache}, R.~R. and {Goldman}, A. and {Perevalov}, V.~I. and {Tashkun}, S.~A. and {Tennyson}, J.},
        title = "{HITEMP, the high-temperature molecular spectroscopic database}",
      journal = {\jqsrt},
         year = 2010,
        month = oct,
       volume = {111},
        pages = {2139-2150},
          doi = {10.1016/j.jqsrt.2010.05.001},
       adsurl = {https://ui.adsabs.harvard.edu/abs/2010JQSRT.111.2139R}
}

@ARTICLE{Barber:2014,
       author = {{Barber}, R.~J. and {Strange}, J.~K. and {Hill}, C. and {Polyansky}, O.~L. and {Mellau}, G. Ch. and {Yurchenko}, S.~N. and {Tennyson}, Jonathan},
        title = "{ExoMol line lists - III. An improved hot rotation-vibration line list for HCN and HNC}",
      journal = {\mnras},
         year = 2014,
        month = jan,
       volume = {437},
       number = {2},
        pages = {1828-1835},
          doi = {10.1093/mnras/stt2011},
archivePrefix = {arXiv},
       eprint = {1311.1328},
 primaryClass = {astro-ph.SR},
       adsurl = {https://ui.adsabs.harvard.edu/abs/2014MNRAS.437.1828B}
}

@ARTICLE{Coles:2019,
       author = {{Coles}, Phillip A. and {Yurchenko}, Sergei N. and {Tennyson}, Jonathan},
        title = "{ExoMol molecular line lists - XXXV. A rotation-vibration line list for hot ammonia}",
      journal = {\mnras},
         year = 2019,
        month = dec,
       volume = {490},
       number = {4},
        pages = {4638-4647},
          doi = {10.1093/mnras/stz2778},
archivePrefix = {arXiv},
       eprint = {1911.10369},
 primaryClass = {astro-ph.SR},
       adsurl = {https://ui.adsabs.harvard.edu/abs/2019MNRAS.490.4638C}
}

@ARTICLE{Leconte:2021,
       author = {{Leconte}, J{\'e}r{\'e}my},
        title = "{Spectral binning of precomputed correlated-k coefficients}",
      journal = {\aap},
         year = 2021,
        month = jan,
       volume = {645},
          eid = {A20},
        pages = {A20},
          doi = {10.1051/0004-6361/202039040},
archivePrefix = {arXiv},
       eprint = {2012.01428},
 primaryClass = {astro-ph.EP},
       adsurl = {https://ui.adsabs.harvard.edu/abs/2021A&A...645A..20L}
}

@ARTICLE{Guillot:2010,
       author = {{Guillot}, T.},
        title = "{On the radiative equilibrium of irradiated planetary atmospheres}",
      journal = {\aap},
         year = 2010,
        month = sep,
       volume = {520},
          eid = {A27},
        pages = {A27},
          doi = {10.1051/0004-6361/200913396},
archivePrefix = {arXiv},
       eprint = {1006.4702},
 primaryClass = {astro-ph.EP},
       adsurl = {https://ui.adsabs.harvard.edu/abs/2010A&A...520A..27G}
}

@article{benneke2019water,
  title={Water vapor and clouds on the habitable-zone sub-Neptune exoplanet K2-18b},
  author={Benneke, Bj{\"o}rn and Wong, Ian and Piaulet, Caroline and Knutson, Heather A and Lothringer, Joshua and Morley, Caroline V and Crossfield, Ian JM and Gao, Peter and Greene, Thomas P and Dressing, Courtney and others},
  journal={The Astrophysical Journal Letters},
  volume={887},
  number={1},
  pages={L14},
  year={2019},
  publisher={The American Astronomical Society}
}

@article{tsiaras2019water,
  title={Water vapour in the atmosphere of the habitable-zone eight-Earth-mass planet K2-18 b},
  author={Tsiaras, Angelos and Waldmann, Ingo P and Tinetti, Giovanna and Tennyson, Jonathan and Yurchenko, Sergey N},
  journal={Nature Astronomy},
  volume={3},
  number={12},
  pages={1086--1091},
  year={2019},
  publisher={Nature Publishing Group UK London}
}

@article{madhusudhan2023carbon,
  title={Carbon-bearing molecules in a possible Hycean atmosphere},
  author={Madhusudhan, Nikku and Sarkar, Subhajit and Constantinou, Savvas and Holmberg, M{\aa}ns and Piette, Anjali AA and Moses, Julianne I},
  journal={The Astrophysical Journal Letters},
  volume={956},
  number={1},
  pages={L13},
  year={2023},
  publisher={IOP Publishing}
}

@article{wogan2024jwst,
  title={JWST observations of K2-18b can be explained by a gas-rich mini-Neptune with no habitable surface},
  author={Wogan, Nicholas F and Batalha, Natasha E and Zahnle, Kevin J and Krissansen-Totton, Joshua and Tsai, Shang-Min and Hu, Renyu},
  journal={The Astrophysical Journal Letters},
  volume={963},
  number={1},
  pages={L7},
  year={2024},
  publisher={The American Astronomical Society}
}

@article{ehrenreich2014near,
  title={Near-infrared transmission spectrum of the warm-Uranus GJ 3470b with the Wide Field Camera-3 on the Hubble Space Telescope},
  author={Ehrenreich, David and Bonfils, Xavier and Lovis, Christophe and Delfosse, Xavier and Forveille, Thierry and Mayor, Michel and Neves, Vasco and Santos, Nuno C and Udry, St{\'e}phane and S{\'e}gransan, Damien},
  journal={Astronomy \& Astrophysics},
  volume={570},
  pages={A89},
  year={2014},
  publisher={EDP Sciences}
}

@article{benneke2019sub,
  title={A sub-Neptune exoplanet with a low-metallicity methane-depleted atmosphere and Mie-scattering clouds},
  author={Benneke, Bj{\"o}rn and Knutson, Heather A and Lothringer, Joshua and Crossfield, Ian JM and Moses, Julianne I and Morley, Caroline and Kreidberg, Laura and Fulton, Benjamin J and Dragomir, Diana and Howard, Andrew W and others},
  journal={Nature Astronomy},
  volume={3},
  number={9},
  pages={813--821},
  year={2019},
  publisher={Nature Publishing Group UK London}
}

@article{berta2012flat,
  title={The flat transmission spectrum of the super-Earth GJ1214b from wide field camera 3 on the Hubble Space Telescope},
  author={Berta, Zachory K and Charbonneau, David and D{\'e}sert, Jean-Michel and Miller-Ricci Kempton, Eliza and McCullough, Peter R and Burke, Christopher J and Fortney, Jonathan J and Irwin, Jonathan and Nutzman, Philip and Homeier, Derek},
  journal={The Astrophysical Journal},
  volume={747},
  number={1},
  pages={35},
  year={2012},
  publisher={The American Astronomical Society}
}

@article{kreidberg2014clouds,
  title={Clouds in the atmosphere of the super-Earth exoplanet GJ 1214b},
  author={Kreidberg, Laura and Bean, Jacob L and D{\'e}sert, Jean-Michel and Benneke, Bj{\"o}rn and Deming, Drake and Stevenson, Kevin B and Seager, Sara and Berta-Thompson, Zachory and Seifahrt, Andreas and Homeier, Derek},
  journal={Nature},
  volume={505},
  number={7481},
  pages={69--72},
  year={2014},
  publisher={Nature Publishing Group UK London}
}

@article{kempton2023reflective,
  title={A reflective, metal-rich atmosphere for GJ 1214b from its JWST phase curve},
  author={Kempton, Eliza M-R and Zhang, Michael and Bean, Jacob L and Steinrueck, Maria E and Piette, Anjali AA and Parmentier, Vivien and Malsky, Isaac and Roman, Michael T and Rauscher, Emily and Gao, Peter and others},
  journal={Nature},
  volume={620},
  number={7972},
  pages={67--71},
  year={2023},
  publisher={Nature Publishing Group UK London}
}

@article{ohno2025possible,
  title={A possible metal-dominated atmosphere below the thick aerosols of GJ 1214 b suggested by its JWST panchromatic transmission spectrum},
  author={Ohno, Kazumasa and Schlawin, Everett and Bell, Taylor J and Murphy, Matthew M and Beatty, Thomas G and Welbanks, Luis and Greene, Thomas P and Fortney, Jonathan J and Parmentier, Vivien and Edelman, Isaac R and others},
  journal={The Astrophysical Journal Letters},
  volume={979},
  number={1},
  pages={L7},
  year={2025},
  publisher={IOP Publishing}
}

@article{schlawin2024possible,
  title={Possible carbon dioxide above the thick aerosols of GJ 1214 b},
  author={Schlawin, Everett and Ohno, Kazumasa and Bell, Taylor J and Murphy, Matthew M and Welbanks, Luis and Beatty, Thomas G and Greene, Thomas P and Fortney, Jonathan J and Parmentier, Vivien and Edelman, Isaac R and others},
  journal={The Astrophysical Journal Letters},
  volume={974},
  number={2},
  pages={L33},
  year={2024},
  publisher={The American Astronomical Society}
}

@article{mikal2023hubble,
  title={Hubble Space Telescope transmission spectroscopy for the temperate sub-Neptune TOI-270 d: A possible hydrogen-rich atmosphere containing water vapor},
  author={Mikal-Evans, Thomas and Madhusudhan, Nikku and Dittmann, Jason and G{\"u}nther, Maximilian N and Welbanks, Luis and Van Eylen, Vincent and Crossfield, Ian JM and Daylan, Tansu and Kreidberg, Laura},
  journal={The Astronomical Journal},
  volume={165},
  number={3},
  pages={84},
  year={2023},
  publisher={The American Astronomical Society}
}

@article{benneke2024jwst,
  title={JWST Reveals CH $ \_4 $, CO $ \_2 $, and H $ \_2 $ O in a Metal-rich Miscible Atmosphere on a Two-Earth-Radius Exoplanet},
  author={Benneke, Bj{\"o}rn and Roy, Pierre-Alexis and Coulombe, Louis-Philippe and Radica, Michael and Piaulet, Caroline and Ahrer, Eva-Maria and Pierrehumbert, Raymond and Krissansen-Totton, Joshua and Schlichting, Hilke E and Hu, Renyu and others},
  journal={arXiv preprint arXiv:2403.03325},
  year={2024}
}

@article{libby2026james,
  title={The James Webb Space Telescope NIRSpec-PRISM Transmission Spectrum of the Super-puff, Kepler-51d},
  author={Libby-Roberts, Jessica E and Bello-Arufe, Aaron and Berta-Thompson, Zachory K and Ca{\~n}as, Caleb I and Chachan, Yayaati and Hu, Renyu and Kawashima, Yui and Murray, Catriona and Ohno, Kazumasa and Tokadjian, Armen and others},
  journal={The Astronomical Journal},
  volume={171},
  number={4},
  pages={221},
  year={2026},
  publisher={The American Astronomical Society}
}

@article{holmberg2024possible,
  title={Possible Hycean conditions in the sub-Neptune TOI-270 d},
  author={Holmberg, M{\aa}ns and Madhusudhan, Nikku},
  journal={Astronomy \& Astrophysics},
  volume={683},
  pages={L2},
  year={2024},
  publisher={EDP Sciences}
}

@article{knutson2014featureless,
  title={A featureless transmission spectrum for the Neptune-mass exoplanet GJ 436b},
  author={Knutson, Heather A and Benneke, Bj{\"o}rn and Deming, Drake and Homeier, Derek},
  journal={Nature},
  volume={505},
  number={7481},
  pages={66--68},
  year={2014},
  publisher={Nature Publishing Group UK London}
}

@article{knutson2014hubble,
  title={Hubble Space Telescope near-IR transmission spectroscopy of the super-Earth HD 97658b},
  author={Knutson, Heather A and Dragomir, Diana and Kreidberg, Laura and Kempton, Eliza M-R and McCullough, PR and Fortney, Jonathan J and Bean, Jacob L and Gillon, Michael and Homeier, Derek and Howard, Andrew W},
  journal={The Astrophysical Journal},
  volume={794},
  number={2},
  pages={155},
  year={2014},
  publisher={The American Astronomical Society}
}

@article{guo2020updated,
  title={Updated parameters and a new transmission spectrum of HD 97658b},
  author={Guo, Xueying and Crossfield, Ian JM and Dragomir, Diana and Kosiarek, Molly R and Lothringer, Joshua and Mikal-Evans, Thomas and Rosenthal, Lee and Benneke, Bjorn and Knutson, Heather A and Dalba, Paul A and others},
  journal={The Astronomical Journal},
  volume={159},
  number={5},
  pages={239},
  year={2020},
  publisher={The American Astronomical Society}
}

@article{mikal2021transmission,
  title={Transmission spectroscopy for the warm sub-Neptune HD 3167c: evidence for molecular absorption and a possible high-metallicity atmosphere},
  author={Mikal-Evans, Thomas and Crossfield, Ian JM and Benneke, Bj{\"o}rn and Kreidberg, Laura and Moses, Julie and Morley, Caroline V and Thorngren, Daniel and Molli{\`e}re, Paul and Hardegree-Ullman, Kevin K and Brewer, John and others},
  journal={The Astronomical Journal},
  volume={161},
  number={1},
  pages={18},
  year={2021},
  publisher={The American Astronomical Society}
}

@article{kreidberg2022tentative,
  title={Tentative evidence for water vapor in the atmosphere of the Neptune-sized exoplanet HD 106315c},
  author={Kreidberg, Laura and Molli{\`e}re, Paul and Crossfield, Ian JM and Thorngren, Daniel P and Kawashima, Yui and Morley, Caroline V and Benneke, Bj{\"o}rn and Mikal-Evans, Thomas and Berardo, David and Kosiarek, Molly R and others},
  journal={The Astronomical Journal},
  volume={164},
  number={4},
  pages={124},
  year={2022},
  publisher={The American Astronomical Society}
}

@article{roy2023water,
  title={Water absorption in the transmission spectrum of the water world candidate GJ 9827 d},
  author={Roy, Pierre-Alexis and Benneke, Bj{\"o}rn and Piaulet, Caroline and Gully-Santiago, Michael A and Crossfield, Ian JM and Morley, Caroline V and Kreidberg, Laura and Mikal-Evans, Thomas and Brande, Jonathan and Delisle, Simon and others},
  journal={The Astrophysical Journal Letters},
  volume={954},
  number={2},
  pages={L52},
  year={2023},
  publisher={The American Astronomical Society}
}

@article{crossfield2020phase,
  title={Phase curves of hot Neptune LTT 9779b suggest a high-metallicity atmosphere},
  author={Crossfield, Ian JM and Dragomir, Diana and Cowan, Nicolas B and Daylan, Tansu and Wong, Ian and Kataria, Tiffany and Deming, Drake and Kreidberg, Laura and Mikal-Evans, Thomas and Gorjian, Varoujan and others},
  journal={The Astrophysical Journal Letters},
  volume={903},
  number={1},
  pages={L7},
  year={2020},
  publisher={The American Astronomical Society}
}

@article{dragomir2020spitzer,
  title={Spitzer reveals evidence of molecular absorption in the atmosphere of the hot Neptune LTT 9779b},
  author={Dragomir, Diana and Crossfield, Ian JM and Benneke, Bj{\"o}rn and Wong, Ian and Daylan, Tansu and Diaz, Matias and Deming, Drake and Molliere, Paul and Kreidberg, Laura and Jenkins, James S and others},
  journal={The Astrophysical Journal Letters},
  volume={903},
  number={1},
  pages={L6},
  year={2020},
  publisher={The American Astronomical Society}
}

@article{edwards2023characterizing,
  title={Characterizing a world within the hot-neptune desert: Transit observations of LTT 9779 b with the hubble space telescope/WFC3},
  author={Edwards, Billy and Changeat, Quentin and Tsiaras, Angelos and Allan, Andrew and Behr, Patrick and Hagey, Simone R and Himes, Michael D and Ma, Sushuang and Stassun, Keivan G and Thomas, Luis and others},
  journal={The Astronomical Journal},
  volume={166},
  number={4},
  pages={158},
  year={2023},
  publisher={The American Astronomical Society}
}

@article{radica2024muted,
  title={Muted features in the JWST NIRISS transmission spectrum of hot Neptune LTT 9779b},
  author={Radica, Michael and Coulombe, Louis-Philippe and Taylor, Jake and Albert, Loic and Allart, Romain and Benneke, Bj{\"o}rn and Cowan, Nicolas B and Dang, Lisa and Lafreni{\`e}re, David and Thorngren, Daniel and others},
  journal={The Astrophysical Journal Letters},
  volume={962},
  number={1},
  pages={L20},
  year={2024},
  publisher={The American Astronomical Society}
}

@article{fraine2014water,
  title={Water vapour absorption in the clear atmosphere of a Neptune-sized exoplanet},
  author={Fraine, Jonathan and Deming, Drake and Benneke, Bjorn and Knutson, Heather and Jord{\'a}n, Andr{\'e}s and Espinoza, N{\'e}stor and Madhusudhan, Nikku and Wilkins, Ashlee and Todorov, Kamen},
  journal={Nature},
  volume={513},
  number={7519},
  pages={526--529},
  year={2014},
  publisher={Nature Publishing Group UK London}
}

@article{chachan2019hubble,
  title={A Hubble PanCET Study of HAT-P-11b: a cloudy Neptune with a low atmospheric metallicity},
  author={Chachan, Yayaati and Knutson, Heather A and Gao, Peter and Kataria, Tiffany and Wong, Ian and Henry, Gregory W and Benneke, Bjorn and Zhang, Michael and Barstow, Joanna and Bean, Jacob L and others},
  journal={The Astronomical Journal},
  volume={158},
  number={6},
  pages={244},
  year={2019},
  publisher={The American Astronomical Society}
}

@article{murphy2025hst,
  title={HST Transmission Spectra of the Hot Neptune HD 219666 b: Detection of Water and the Challenge of Constraining Both Water and Methane},
  author={Murphy, Matthew M and Beatty, Thomas G and Welbanks, Luis and Fu, Guangwei},
  journal={The Astronomical Journal},
  volume={169},
  number={6},
  pages={286},
  year={2025},
  publisher={The American Astronomical Society}
}

@article{hu2025water,
  title={A water-rich interior in the temperate sub-Neptune K2-18 b revealed by JWST},
  author={Hu, Renyu and Bello-Arufe, Aaron and Tokadjian, Armen and Yang, Jeehyun and Damiano, Mario and Roy, Pierre-Alexis and Coulombe, Louis-Philippe and Madhusudhan, Nikku and Constantinou, Savvas and Benneke, Bj{\"o}rn},
  journal={arXiv preprint arXiv:2507.12622},
  year={2025}
}

@article{blain20211d,
  title={1D atmospheric study of the temperate sub-Neptune K2-18b},
  author={Blain, Doriann and Charnay, Benjamin and B{\'e}zard, Bruno},
  journal={Astronomy \& Astrophysics},
  volume={646},
  pages={A15},
  year={2021},
  publisher={EDP Sciences}
}

@article{morley2015thermal,
  title={Thermal emission and reflected light spectra of super Earths with flat transmission spectra},
  author={Morley, Caroline V and Fortney, Jonathan J and Marley, Mark S and Zahnle, Kevin and Line, Michael and Kempton, Eliza and Lewis, Nikole and Cahoy, Kerri},
  journal={The Astrophysical Journal},
  volume={815},
  number={2},
  pages={110},
  year={2015},
  publisher={The American Astronomical Society}
}

@article{fortney2008synthetic,
  title={Synthetic spectra and colors of young giant planet atmospheres: effects of initial conditions and atmospheric metallicity},
  author={Fortney, Jonathan J and Marley, Mark S and Saumon, Didier and Lodders, Katharina},
  journal={The Astrophysical Journal},
  volume={683},
  number={2},
  pages={1104--1116},
  year={2008}
}

@article{marley1999reflected,
  title={Reflected spectra and albedos of extrasolar giant planets. I. Clear and cloudy atmospheres},
  author={Marley, Mark S and Gelino, Christopher and Stephens, Denise and Lunine, Jonathan I and Freedman, Richard},
  journal={The Astrophysical Journal},
  volume={513},
  number={2},
  pages={879--893},
  year={1999}
}

@article{husser2013new,
  title={A new extensive library of PHOENIX stellar atmospheres and synthetic spectra},
  author={Husser, T-O and Wende-von Berg, Sebastian and Dreizler, Stefan and Homeier, Derek and Reiners, Ansgar and Barman, Travis and Hauschildt, Peter H},
  journal={Astronomy \& Astrophysics},
  volume={553},
  pages={A6},
  year={2013},
  publisher={EDP Sciences}
}

@article{ackerman2001precipitating,
  title={Precipitating condensation clouds in substellar atmospheres},
  author={Ackerman, Andrew S and Marley, Mark S},
  journal={The Astrophysical Journal},
  volume={556},
  number={2},
  pages={872--884},
  year={2001}
}

@article{foreman2013emcee,
  title={emcee: the MCMC hammer},
  author={Foreman-Mackey, Daniel and Hogg, David W and Lang, Dustin and Goodman, Jonathan},
  journal={Publications of the Astronomical Society of the Pacific},
  volume={125},
  number={925},
  pages={306},
  year={2013},
  publisher={IOP Publishing}
}

@article{brande2024clouds,
  title={Clouds and clarity: Revisiting atmospheric feature trends in Neptune-size exoplanets},
  author={Brande, Jonathan and Crossfield, Ian JM and Kreidberg, Laura and Morley, Caroline V and Barman, Travis and Benneke, Bj{\"o}rn and Christiansen, Jessie L and Dragomir, Diana and Fortney, Jonathan J and Greene, Thomas P and others},
  journal={The Astrophysical Journal Letters},
  volume={961},
  number={1},
  pages={L23},
  year={2024},
  publisher={The American Astronomical Society}
}

@ARTICLE{Welbanks2024,
       author = {{Welbanks}, Luis and {Bell}, Taylor J. and {Beatty}, Thomas G. and {Line}, Michael R. and {Ohno}, Kazumasa and {Fortney}, Jonathan J. and {Schlawin}, Everett and {Greene}, Thomas P. and {Rauscher}, Emily and {McGill}, Peter and {Murphy}, Matthew and {Parmentier}, Vivien and {Tang}, Yao and {Edelman}, Isaac and {Mukherjee}, Sagnick and {Wiser}, Lindsey S. and {Lagage}, Pierre-Olivier and {Dyrek}, Achr{\`e}ne and {Arnold}, Kenneth E.},
        title = "{A high internal heat flux and large core in a warm Neptune exoplanet}",
      journal = {\nat},
         year = 2024,
        month = jun,
       volume = {630},
       number = {8018},
        pages = {836-840},
          doi = {10.1038/s41586-024-07514-w},
archivePrefix = {arXiv},
       eprint = {2405.11018},
 primaryClass = {astro-ph.EP},
       adsurl = {https://ui.adsabs.harvard.edu/abs/2024Natur.630..836W}
}

@article{beatty2024sulfur,
  title={Sulfur dioxide and other molecular species in the atmosphere of the sub-Neptune GJ 3470 b},
  author={Beatty, Thomas G and Welbanks, Luis and Schlawin, Everett and Bell, Taylor J and Line, Michael R and Murphy, Matthew and Edelman, Isaac and Greene, Thomas P and Fortney, Jonathan J and Henry, Gregory W and others},
  journal={The Astrophysical Journal Letters},
  volume={970},
  number={1},
  pages={L10},
  year={2024},
  publisher={The American Astronomical Society}
}

@article{madhusudhan2025exploring,
  title={Exploring the sub-Neptune frontier with JWST},
  author={Madhusudhan, Nikku and Holmberg, M{\aa}ns and Constantinou, Savvas and Cooke, Gregory J},
  journal={Proceedings of the National Academy of Sciences},
  volume={122},
  number={39},
  pages={e2416194122},
  year={2025},
  publisher={National Academy of Sciences}
}

@article{knapp2004near,
  title={Near-infrared photometry and spectroscopy of L and T dwarfs: The effects of temperature, clouds, and gravity},
  author={Knapp, GR and Leggett, Sandy K and Fan, X and Marley, MS and Geballe, TR and Golimowski, DA and Finkbeiner, Douglas and Gunn, JE and Hennawi, J and Ivezi{\'c}, Z and others},
  journal={The Astronomical Journal},
  volume={127},
  number={6},
  pages={3553--3578},
  year={2004}
}

@article{tsuji2003transition,
  title={Transition from l to t dwarfs on the color-magnitude diagram},
  author={Tsuji, Takashi and Nakajima, Tadashi},
  journal={The Astrophysical Journal Letters},
  volume={585},
  number={2},
  pages={L151--L154},
  year={2003}
}

@article{buenzli2014brown,
  title={Brown dwarf photospheres are patchy: a hubble space telescope near-infrared spectroscopic survey finds frequent low-level variability},
  author={Buenzli, Esther and Apai, D{\'a}niel and Radigan, Jacqueline and Reid, I Neill and Flateau, Davin},
  journal={The Astrophysical Journal},
  volume={782},
  number={2},
  pages={77},
  year={2014},
  publisher={The American Astronomical Society}
}

@ARTICLE{pacman,
       author = {{Zieba}, Sebastian and {Kreidberg}, Laura},
        title = "{PACMAN: A pipeline to reduce and analyze Hubble Wide Field Camera 3 IR Grism data}",
      journal = {The Journal of Open Source Software},
         year = 2022,
        month = dec,
       volume = {7},
       number = {80},
          eid = {4838},
        pages = {4838},
          doi = {10.21105/joss.04838},
archivePrefix = {arXiv},
       eprint = {2212.11421},
 primaryClass = {astro-ph.IM},
       adsurl = {https://ui.adsabs.harvard.edu/abs/2022JOSS....7.4838Z}
}

@book{irwin09,
  title={Giant planets of our solar system: atmospheres, composition, and structure},
  author={Irwin, Patrick},
  year={2009},
  publisher={Springer Science \& Business Media}
}

@ARTICLE{exoplanetarchive,
       author = {{Christiansen}, Jessie L. and {McElroy}, Douglas L. and {Harbut}, Marcy and {Ciardi}, David R. and {Crane}, Megan and {Good}, John and {Hardegree-Ullman}, Kevin K. and {Kesseli}, Aurora Y. and {Lund}, Michael B. and {Lynn}, Meca and {Muthiar}, Ananda and {Nilsson}, Ricky and {Oluyide}, Toba and {Papin}, Michael and {Rivera}, Amalia and {Swain}, Melanie and {Susemiehl}, Nicholas D. and {Tam}, Raymond and {van Eyken}, Julian and {Beichman}, Charles},
        title = "{The NASA Exoplanet Archive and Exoplanet Follow-up Observing Program: Data, Tools, and Usage}",
      journal = {PSJ},
         year = 2025,
        month = aug,
       volume = {6},
       number = {8},
          eid = {186},
        pages = {186},
          doi = {10.3847/PSJ/ade3c2},
archivePrefix = {arXiv},
       eprint = {2506.03299},
 primaryClass = {astro-ph.EP},
       adsurl = {https://ui.adsabs.harvard.edu/abs/2025PSJ.....6..186C}
}

@article{oberg2011effects,
  title={The effects of snowlines on C/O in planetary atmospheres},
  author={{\"O}berg, Karin I and Murray-Clay, Ruth and Bergin, Edwin A},
  journal={The Astrophysical Journal Letters},
  volume={743},
  number={1},
  pages={L16},
  year={2011},
  publisher={The American Astronomical Society}
}

@ARTICLE{crossfield:2017b,
   author = {{Crossfield}, I.~J.~M. and {Kreidberg}, L.},
    title = "{Trends in Atmospheric Properties of Neptune-size Exoplanets}",
  journal = {\aj},
archivePrefix = "arXiv",
   eprint = {1708.00016},
 primaryClass = "astro-ph.EP",
     year = 2017,
    month = dec,
   volume = 154,
      eid = {261},
    pages = {261},
      doi = {10.3847/1538-3881/aa9279},
   adsurl = {http://adsabs.harvard.edu/abs/2017AJ....154..261C}
}

@article{yu2021haze,
  title={Haze evolution in temperate exoplanet atmospheres through surface energy measurements},
  author={Yu, Xinting and He, Chao and Zhang, Xi and H{\"o}rst, Sarah M and Dymont, Austin H and McGuiggan, Patricia and Moses, Julianne I and Lewis, Nikole K and Fortney, Jonathan J and Gao, Peter and others},
  journal={Nature Astronomy},
  volume={5},
  number={8},
  pages={822--831},
  year={2021},
  publisher={Nature Publishing Group UK London}
}

@article{bean2021nature,
  title={The nature and origins of sub-Neptune size planets},
  author={Bean, Jacob L and Raymond, Sean N and Owen, James E},
  journal={Journal of Geophysical Research: Planets},
  volume={126},
  number={1},
  pages={e2020JE006639},
  year={2021},
  publisher={Wiley Online Library}
}

@article{libby2020featureless,
  title={The featureless transmission spectra of two super-puff planets},
  author={Libby-Roberts, Jessica E and Berta-Thompson, Zachory K and D{\'e}sert, Jean-Michel and Masuda, Kento and Morley, Caroline V and Lopez, Eric D and Deck, Katherine M and Fabrycky, Daniel and Fortney, Jonathan J and Line, Michael R and others},
  journal={The Astronomical Journal},
  volume={159},
  number={2},
  pages={57},
  year={2020},
  publisher={The American Astronomical Society}
}

@article{wallack2024jwst,
  title={JWST COMPASS: A NIRSpec/G395H transmission spectrum of the sub-Neptune TOI-836c},
  author={Wallack, Nicole L and Batalha, Natasha E and Alderson, Lili and Scarsdale, Nicholas and Adams Redai, Jea I and Aguichine, Artyom and Alam, Munazza K and Gao, Peter and Wolfgang, Angie and Batalha, Natalie M and others},
  journal={The Astronomical Journal},
  volume={168},
  number={2},
  pages={77},
  year={2024},
  publisher={The American Astronomical Society}
}

@article{gordon2026jwst,
  title={JWST COMPASS: Insights into the Systematic Noise Properties of NIRSpec/G395H from a Uniform Reanalysis of Seven Transmission Spectra},
  author={Gordon, Tyler A and Batalha, Natalie M and Batalha, Natasha E and Aguichine, Artyom and Gagnebin, Anna and Kirk, James and L{\'o}pez-Morales, Mercedes and Meech, Annabella and Scarsdale, Nicholas and Teske, Johanna and others},
  journal={The Astronomical Journal},
  volume={171},
  number={3},
  pages={178},
  year={2026},
  publisher={The American Astronomical Society}
}

@article{wallack2026jwst,
  title={JWST COMPASS: NIRSpec/G395H Transmission Observations of the Sub-Neptune HD 15337 c},
  author={Wallack, Nicole L and Gao, Peter and Greklek-McKeon, Michael and Meech, Annabella and Aguichine, Artyom and Alam, Munazza K and Alderson, Lili and Batalha, Natasha E and Batalha, Natalie M and Gagnebin, Anna and others},
  journal={The Astronomical Journal},
  volume={171},
  number={3},
  pages={180},
  year={2026},
  publisher={The American Astronomical Society}
}

@article{meech2026jwst,
  title={JWST COMPASS: A NIRSpec G395H Transmission Spectrum of Radius Valley Dweller TOI-260 b},
  author={Meech, Annabella and Gao, Peter and Wallack, Nicole L and L{\'o}pez-Morales, Mercedes and Oddo, Dominic and Teske, Johanna and Dragomir, Diana and Wolfgang, Angie and Wogan, Nicholas and Wakeford, Hannah R and others},
  journal={The Astronomical Journal},
  volume={171},
  number={5},
  pages={274},
  year={2026},
  publisher={The American Astronomical Society}
}

@article{sarkar2026transmission,
  title={Transmission spectrum of the benchmark temperate exo-Neptune TOI-1231 b},
  author={Sarkar, Subhajit and Madhusudhan, Nikku and Pica-Ciamarra, Lorenzo and Holmberg, M{\aa}ns and Rigby, Frances E and Moses, Julianne I and Mealing, Megan},
  journal={The Astrophysical Journal Letters},
  volume={1005},
  number={2},
  pages={L52},
  year={2026},
  publisher={The American Astronomical Society}
}

@article{rigby2025jwst,
  title={A jwst transmission spectrum of the temperate sub-neptune toi-732 c},
  author={Rigby, Frances E and Madhusudhan, Nikku and Sarkar, Subhajit and Pica-Ciamarra, Lorenzo and Holmberg, M{\aa}ns and Moses, Julianne I},
  journal={The Astrophysical Journal Letters},
  volume={995},
  number={2},
  pages={L70},
  year={2025},
  publisher={The American Astronomical Society}
}

@article{zeng2019growth,
  title={Growth model interpretation of planet size distribution},
  author={Zeng, Li and Jacobsen, Stein B and Sasselov, Dimitar D and Petaev, Michail I and Vanderburg, Andrew and Lopez-Morales, Mercedes and Perez-Mercader, Juan and Mattsson, Thomas R and Li, Gongjie and Heising, Matthew Z and others},
  journal={Proceedings of the National Academy of Sciences},
  volume={116},
  number={20},
  pages={9723--9728},
  year={2019},
  publisher={National Academy of Sciences}
}

@ARTICLE{aguichine2021,
       author = {{Aguichine}, Artyom and {Mousis}, Olivier and {Deleuil}, Magali and {Marcq}, Emmanuel},
        title = "{Mass-Radius Relationships for Irradiated Ocean Planets}",
      journal = {\apj},
         year = 2021,
        month = jun,
       volume = {914},
       number = {2},
          eid = {84},
        pages = {84},
          doi = {10.3847/1538-4357/abfa99},
archivePrefix = {arXiv},
       eprint = {2105.01102},
 primaryClass = {astro-ph.EP},
       adsurl = {https://ui.adsabs.harvard.edu/abs/2021ApJ...914...84A}
}

@article{wordsworth2013water,
  title={Water loss from terrestrial planets with CO2-rich atmospheres},
  author={Wordsworth, Robin D and Pierrehumbert, Raymond T},
  journal={The Astrophysical Journal},
  volume={778},
  number={2},
  pages={154},
  year={2013},
  publisher={The American Astronomical Society}
}

@article{parmentier20133d,
  title={3D mixing in hot Jupiters atmospheres: I. Application to the day/night cold trap in HD 209458b},
  author={Parmentier, Vivien and Showman, Adam P and Lian, Yuan},
  journal={Astronomy \& Astrophysics},
  volume={558},
  pages={A91},
  year={2013},
  publisher={EDP Sciences}
}

@article{leconte20133d,
  title={3D climate modeling of close-in land planets: circulation patterns, climate moist bistability, and habitability},
  author={Leconte, J{\'e}r{\'e}my and Forget, Francois and Charnay, Benjamin and Wordsworth, Robin and Selsis, Franck and Millour, Ehouarn and Spiga, Aymeric},
  journal={Astronomy \& Astrophysics},
  volume={554},
  pages={A69},
  year={2013},
  publisher={EDP Sciences}
}

@article{helling2021cloud,
  title={Cloud property trends in hot and ultra-hot giant gas planets (WASP-43b, WASP-103b, WASP-121b, HAT-P-7b, and WASP-18b)},
  author={Helling, Ch and Lewis, D and Samra, D and Carone, L and Graham, V and Herbort, O and Chubb, KL and Min, M and Waters, R and Parmentier, V and others},
  journal={Astronomy \& Astrophysics},
  volume={649},
  pages={A44},
  year={2021},
  publisher={EDP Sciences}
}

\end{document}